\documentclass[]{aa}  

\usepackage{graphicx}
\usepackage{txfonts}
\usepackage{hyperref}
\hypersetup
{
    colorlinks= true,
    linkcolor = blue,
    filecolor = blue,      
    urlcolor  = blue,
    citecolor = blue
}
\usepackage{cleveref}
\usepackage{outlines}
\usepackage{xcolor}
\usepackage{CJKutf8}
\usepackage{booktabs}
\usepackage{subfigure}
\usepackage{rotating} 
\usepackage{colortbl} 
\usepackage{xcolor}   
\usepackage{multirow} 
\usepackage{lipsum}
\usepackage{lastpage}
\usepackage{makecell}
\usepackage{chemformula}
\def\R#1{\textsc{Rate#1}}
\newcommand{\crir}[2][1.30]{#1\times10^{#2}\,\mathrm{s}^{-1}}

\newcommand{\cm}{\mathrm{cm}^{-3}}
\newcommand{\dens}[1]{10^{#1}\,\mathrm{cm}^{-3}}
\newcommand{\kms}[1]{#1\,\mathrm{km\,s}^{-1}}
\newcommand{\vs}{v_\mathrm{s}}
\newcommand{\av}{A_\mathrm{V}}
\begin{document}

   \title{The curious case of \ch{HCO+}}

   \subtitle{Extreme abundances under extreme conditions}

\author{
         Katarzyna~M.~Dutkowska \inst{\ref{STRW}}\corrauth{dutkowska@strw.leidenuniv.nl}
    \and Bin Jia(\begin{CJK*}{UTF8}{gbsn}贾彬\end{CJK*}) \inst{\ref{STRW}}
    \and Serena Viti\inst{\ref{STRW},\ref{TRA},\ref{UCL}}
    \and Tobias~M.~Dijkhuis \inst{\ref{STRW},\ref{LIC},\ref{IMM}}
    \and Gijs Vermari\"en \inst{\ref{STRW}}
}
\institute{
             Leiden Observatory, Leiden University, P.O. Box 9513, 2300 RA Leiden, The Netherlands\label{STRW}
        \and Transdisciplinary Research Area (TRA) ‘Matter’/Argelander-Institut für Astronomie, University of Bonn, Bonn, Germany\label{TRA}
        \and Department of Physics and Astronomy, University College London, Gower Street, London, UK\label{UCL}
        \and Leiden Institute of Chemistry, Leiden University, P.O. box 9502, 2300 RA Leiden, The Netherlands\label{LIC}
        \and Institute for Molecules and Materials, Radboud University, 6525 AJ Nijmegen, The Netherlands \label{IMM} 
}

   \date{Received XYZ; accepted XYZ}

 
  \abstract
   {\ch{HCO+} is widely observed in both Galactic and extragalactic environments and typically exhibits abundances of $10^{-9}-10^{-8}$. However, recent modeling studies suggest that in environments exposed to elevated cosmic-ray ionization rates and strong thermal or mechanical processing its abundance may increase by several orders of magnitude.}
   {To interpret these predictions, we need to understand the physical conditions that produce extreme \ch{HCO+} abundances and the chemical pathways that drive these enhancements.}
   {We used UCLCHEM, a gas-grain chemical code, to model the chemistry of \ch{HCO+} in dense molecular, protostellar, and shocked gas under elevated cosmic-ray ionization rates ($\zeta \ge 10^{-15}\,\mathrm{s^{-1}}$).}
   {Extreme \ch{HCO+} enhancements leading to $X(\ch{HCO+}) \gtrsim 10^{-4}$ occur only under specific combinations of temperature, density, and cosmic-ray ionization rate, primarily in protostellar and shocked gas. Increasing density generally suppresses the peak \ch{HCO+} abundance, requiring higher ionization rates to produce comparable enhancements. More importantly, the extreme enhancements seem to be very dependent on the chemical network used (in our case UMIST12 versus UMIST22, with the latter leading to extreme abundances). These differences among networks arise from the removal of   the destruction pathway of \ch{HCO+}: \ch{C + HCO+ -> CO + CH+}, and propagate to several other species including \ch{N2H+}, \ch{H2O}, and \ch{H3O+}.}
   {}
\maketitle
\nolinenumbers
%

\section{Introduction}

Cosmic rays are among the primary drivers of interstellar chemistry. By penetrating both diffuse and dense gas, they regulate the ionization balance and initiate gas-phase chemistry through the ionization of \ch{H2} \citep{herbst1973,watson1973}. They affect a wide range of molecules and therefore impact the overall physical and chemical state of the interstellar medium (ISM).

Among the species strongly influenced by cosmic rays is \ch{HCO+}. Its abundance depends on the electron fraction and the CO reservoir, and its chemistry is closely linked to that of \ch{H3+}, which is an established tracer of the cosmic-ray ionization rate (CRIR, $\zeta$) \citep[e.g.,][]{geballe1996,indriolo2007,oka2019}. As a result, \ch{HCO+} is both a key molecule in interstellar chemistry and a useful probe of the column density of molecular hydrogen \citep{gerin2019,liszt2023}, as well as an indirect tracer of the CRIR \citep{vandertak2000,luo2023}.

Given its astrochemical importance, \ch{HCO+} has been widely observed in a broad range of interstellar environments, where its abundances provide valuable constraints on physical conditions. Since its discovery by \citet{klemperer1970}, it has been detected throughout the ISM, in both Galactic and extragalactic regimes \citep[e.g.,][]{lucas1996,garcia-burillo2006,mills2013}. Typical abundances are found to range from $\sim10^{-9}$ to $\sim10^{-8}$ \citep[e.g.,][]{falgarone2006,podio2014,riquelme2018}.

In star-forming regions, abundances of \ch{HCO+} exceeding $\sim10^{-8}$ have been observed, and this excess has been proposed to arise from outflow activity and the associated far-ultraviolet radiation field \citep[e.g.,][]{rawlings2001,rawlings2004,viti2002,arce2006}. \ch{HCO+} abundances can also increase with higher CRIR \citep[e.g.,][]{albertsson2018,tu2024}. 

Recently, \citet[][D25]{dutkowska2025} found that under Galactic Center conditions, where cosmic-ray fluxes are elevated \citep[e.g.,][]{indriolo2015,padovani2020,padovani2022}, \ch{HCO+} abundances can reach exceptionally high values, up to $\sim10^{-4}$ (relative to hydrogen nuclei) for $\zeta \gtrsim 10^{-14}\,\mathrm{s}^{-1}$. This result was obtained in chemical models using the most recent version (\R{22}) of the UMIST Database for Astrochemistry \citep[UDfA; ][]{umist2022}, while models with identical physical conditions but using the previous release \citep[\R{12};][]{umist2012} predicted substantially lower values.

In this work, we investigate the chemical reactions and environments that can lead to very high abundances of \ch{HCO+}. We also discuss potential caveats that should be considered when interpreting these results. The paper is organized as follows. In Section \ref{sect:methodology}, we detail the modeling setup, and in Section \ref{sect:results} we present the main results and outline considerations for future studies. Finally, Section \ref{sect:conclusions} summarizes the implications and differences in \ch{HCO+} chemistry between the two chemical networks.

\section{Methodology}
\label{sect:methodology}

We modeled the chemistry of \ch{HCO+} using \texttt{UCLCHEM}\footnote{https://uclchem.github.io/ version \texttt{v3.5.3}} \citep{uclchem3.0, uclchem4.0}, an open-source, time-dependent gas-grain chemical code. To assess the impact of updated reaction networks, we employed the two most recent versions of the UMIST Database for Astrochemistry \citep[UDfA;][]{umist2012,umist2022}, \R{12} and \R{22}. For the purpose of this work, we use the default \texttt{UCLCHEM} species and ice reactions (the assumed initial abundances can be found in Tab. \ref{tab:elemental_abundances}). Throughout this work, abundances are expressed relative to hydrogen nuclei, where $n_{\rm H}=n(\ch{H})+2n(\ch{H2})$.

\begin{table}
\centering
\caption{Initial gas-phase elemental abundances used in the UCLCHEM models.}
\begin{tabular}{cc}
\hline\hline
\addlinespace[0.1cm]
Element & Abundance (relative to H nuclei$^{a}$) \\
\addlinespace[0.06cm]
\hline
\addlinespace[0.1cm]
He & $1.00\times10^{-1}$ \\
C  & $1.77\times10^{-4}$ \\
O  & $3.34\times10^{-4}$ \\
N  & $6.18\times10^{-5}$ \\
S  & $3.51\times10^{-6}$ \\
Mg & $2.25\times10^{-6}$ \\
Si$^{b}$ & $1.78\times10^{-6}$ \\
Cl & $3.39\times10^{-8}$ \\
P  & $7.78\times10^{-8}$ \\
Fe & $2.01\times10^{-7}$ \\
F  & $3.60\times10^{-8}$ \\
\hline
\end{tabular}
\label{tab:elemental_abundances}
\tablefoot{$^{a}$The total elemental abundance of H is unity by definition and is therefore omitted from the table. The models are initialized with 50\% of the hydrogen nuclei in atomic form and 50\% in \ch{H2}. $^{b}$For the protostellar gas models, the initial abundance of Si was depleted by a factor of 100 following D25.}
\end{table}

\begin{figure}[t!]
    \centering
    \includegraphics[width=1\linewidth]{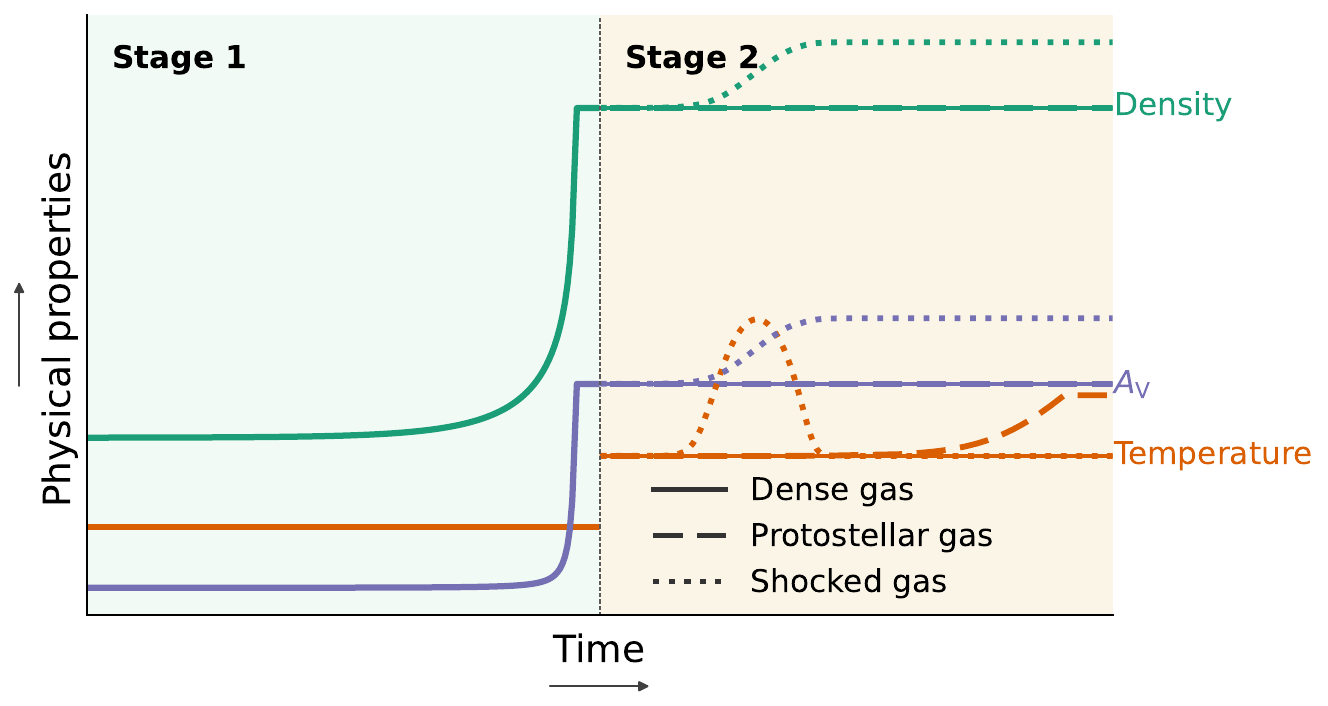}
    \caption{Schematic overview of the physical evolution in the UCLCHEM models presented in this work. During Stage~1, i.e., the isothermal collapse phase, the gas density increases together with the corresponding visual extinction, $\av$, while the temperature remains fixed at 10~K. At the beginning of Stage~2, the initial gas temperature is adjusted to account for cosmic-ray heating, as described in Sect.\ref{sect:methodology}. The subsequent evolution depends on the physical scenario. In the dense molecular gas models, all physical properties remain constant. In the protostellar models, the temperature increases gradually to a predefined maximum value, while the density and $\av$ remain fixed. In the shock models, the gas is compressed, increasing both the density and the corresponding $\av$, while the temperature follows a profile determined by the pre-shock density and shock velocity.}
    \label{fig:schematic}
\end{figure}

Our study considered a broader set of physical structures than D25 to identify environments where extreme \ch{HCO+} abundance enhancements could occur. We set out to test whether such abundances are confined to specific physical structures and conditions, or whether a more cautious interpretation is required. To this end, in addition to protostellar and shocked gas, we explored the behavior of \ch{HCO+} in dense molecular gas. All models presented in this work are single-point (0D), time-dependent models that follow the chemical evolution of a single parcel of gas. The physical evolution of each scenario is summarized schematically in Fig.~\ref{fig:schematic}.

Based on D25, the preferential CRIR range for extreme \ch{HCO+} enhancement lies between $1.3\times10^{-15}$ and $10^{-13}\,\mathrm{s}^{-1}$. We restricted our study to this range, as \citet{bayet2011} showed that \ch{HCO+} gets destroyed at higher CRIR values ($\zeta > 10^{-13}\,\mathrm{s}^{-1}$). We sampled this range logarithmically, selecting five points per order of magnitude to capture possible nonlinear behavior of \ch{HCO+} abundances with increasing CRIR. 

During the formation of a molecular cloud (the ``collapse'' stage, referred to as Stage 1 in \texttt{UCLCHEM}), we used standard CRIR and interstellar UV field strength ($G_0$) to let the chemistry evolve under quiescent conditions. This is in contrast to the scenarios considered in D25, which specifically focused on extreme regions. For the CRIR, we adopted a reference Galactic value of $\crir{-17}$, commonly used in astrochemical modeling of dense molecular clouds \citep[e.g.,][]{caselli1998}, and $G_0=1$~Habing for a standard interstellar radiation field \citep{habing1968}. We also assumed typical initial gas and dust temperatures of 10~K. 

Each cloud isothermally collapsed in free-fall from an initial density of 100~$\cm$ and an initial $\av$ of 2~mag to a final, predefined density, depending on the structure being modeled. For molecular gas evolving into pre-shock regions, the chemistry was allowed to evolve for 1~Myr after reaching the final density. For protostellar and dense molecular gas this modeling stage ended at the completion of collapse.

After the cloud formation stage, we model the objects that constitute the focus of this study. Each object is defined by its characteristic density and visual extinction rather than by a physical size, and is evolved for 1~Myr. The molecular gas cases, including the pre-shock medium, are characterized by $\av \approx 10$ mag, while the protostellar models adopt $\av \geq 100$ mag. The full setup of the Stage 2 models is summarized in Table \ref{tab:models}. 

To model dense molecular gas in Stage 2, we adopt static cloud models in which all physical properties remain constant throughout the evolution. For protostellar and shocked gas, we use protostellar core and C-shock models, respectively. In the protostellar models, the heating profile follows a warm-up around a $10~\mathrm{M}_\odot$ source. For the C-shock models, we explore higher velocities than those considered in D25 and therefore set the magnetic parameter $bm_0=3$, to maintain magnetically supported shocks under these conditions. For a detailed description of the treatment of physical and chemical processes in UCLCHEM, we refer the reader to \cite{uclchem4.0}.

Since all Stage 2 models adopt high CRIRs and the thermal balance is not calculated self-consistently, gas and dust temperatures were prescribed based on the dense cloud PDR models of \citet{bisbas2021}. For $\zeta \leq \crir[1]{-14}$, temperatures are obtained by piecewise-linear interpolation in $\log_{10}(\zeta)$, and by linear extrapolation at higher values, extending up to $\sim$1.1~dex beyond the grid of \citet{bisbas2021}. This yields values in the range $41-77$~K. These temperatures were adopted for the dense molecular gas models and as the initial temperatures of the protostellar and shocked gas models.

\begin{table*}[t!]
\centering
\caption{Physical parameters adopted in the Stage~2 models.}
\begin{tabular}{l c lc}
\hline\hline
\addlinespace[0.1cm]
Property & Unit & Values & Sampling\\
\addlinespace[0.06cm]
\hline
\addlinespace[0.1cm]
\multicolumn{4}{l}{\textbf{Global parameters}} \\
$G_\mathrm{0}$              & Habing       & $10^{2},\,10^{3}$ & -   \\
$\zeta$                     & $\crir{-17}$ & $10^{2}-10^{4}$   & log \\
$t_\mathrm{stage}$          & yr           & $10^6$            & - \\
\addlinespace[0.1cm]
\hline
\addlinespace[0.1cm]
\multicolumn{4}{l}{\textbf{Dense molecular gas}} \\
Density                     & $\cm$        & $10^3 - 10^6$       & log \\
Temperature                 & K            & $41.31 - 76.51^{a}$ & - \\
$\av$                       & mag          & 10                  & - \\
\addlinespace[0.1cm]
\hline
\addlinespace[0.1cm]
\multicolumn{4}{l}{\textbf{Protostellar gas}} \\
Density                      & $\cm$       & $10^6-10^8$    & log    \\
Temperature$_\mathrm{max}$   & K           & 300--500       & linear \\
$\av$                        & mag         & $\gtrsim 100$  & -      \\
\addlinespace[0.1cm]
\hline
\addlinespace[0.1cm]
\multicolumn{4}{l}{\textbf{Shocked gas}} \\
Density$_\mathrm{pre-shock}$ & $\cm$       & $10^3-10^5$   & log    \\
Temperature                  & K           & $41.31-6631^{b}$ & -      \\
$A_\mathrm{V,\,pre-shock}$   & mag         & 10            & -      \\
$\vs$                        & $\kms{}$    & $5-60^{c}$    & linear \\
\addlinespace[0.15cm]
\hline
\end{tabular}
\label{tab:models}
\tablefoot{
$^{a}$Temperatures of dense molecular gas are based on dense cloud models from \citet{bisbas2021}. $^{b}$Temperatures in shocked gas vary from the pre-shocked temperatures determined by $\zeta$ to peak gas temperatures determined by the shock properties. $^{c}$The grid is sampled from 10 to $\kms{60}$ in uniform $\kms{10}$ steps; a single additional model at $\kms{5}$ was included to probe chemistry of very low–velocity shocks.
}
\end{table*}

\section{Results \& discussion}
\label{sect:results}

We examined the chemistry of \ch{HCO+} across a broad range of astrophysical environments. To capture the reactions responsible for strong \ch{HCO+} enhancement, all models were computed using two gas-phase chemical networks, \R{12} and \R{22}, based on the discrepancies reported by D25. Across the explored physical conditions, the extremely high abundance of \ch{HCO+} is found to be primarily regulated by the reaction of \ch{HCO+} with atomic carbon. Additionally, for the extinction values considered here, the radiation field strength $G_0$ has negligible impact on the abundances of the studied molecules, and we therefore do not discuss its effect further. In the following sections we examine the behavior of \ch{HCO+} and chemical differences between the two networks in more detail.

\subsection{Dense molecular gas}

\begin{figure}[t!]
    \centering
    \includegraphics[width=1\linewidth]{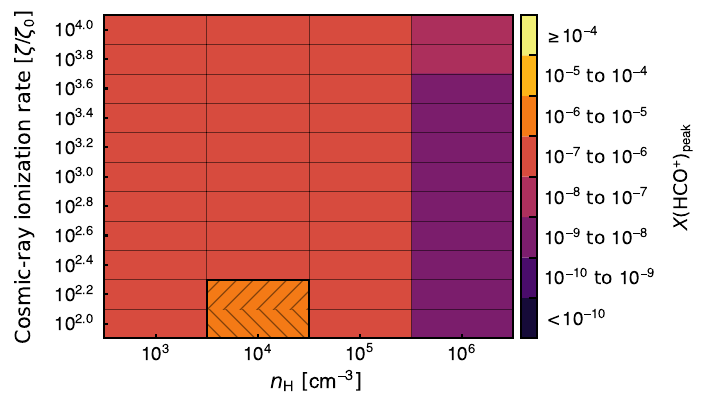}
    \caption{Heatmap of peak \ch{HCO+} abundances for the \R{22} network. Black contours mark regions where $X(\ch{HCO+}) > 10^{-6}$. Forward hatching ($/$) indicates the global maximum obtained with \R{12}, and backward hatching (\textbackslash) indicates the global maximum obtained with \R{22}. For the dense molecular gas, no model exhibits $\Delta_\mathrm{R22/R12} \geq 1$ dex when evaluated at the \R{22} peak abundance.}
    \label{fig:dense_gas_heatmap}
\end{figure}

Dense gas is ubiquitous in the ISM and represents an important component of the molecular reservoir in galaxies. Therefore, even though its physical conditions differ from those associated with the high \ch{HCO+} abundances reported by D25, we examine whether its chemical evolution is similarly affected by elevated CRIRs.

We find that the behavior of \ch{HCO+} is sensitive to both the CRIR $\zeta$ and the gas density $n_{\mathrm{H}}$. At the lowest density, $n_\mathrm{H}=\dens{3}$, the abundance decreases with time and in the case of the highest ionization models, it drops by more than 3~dex within the first $10^4$~yr (see Fig. \ref{fig:DMG_all}). On the other hand, the highest density models, $n_\mathrm{H} = \dens{6}$, tend to yield the lowest \ch{HCO+} abundances among all considered clouds.

The peak abundance among the \R{22} models, $X(\ch{HCO+})_\mathrm{peak} = 1.19 \times 10^{-6}$, is reached for $\zeta = \crir{-15}$, $G_0 = 10^3$~Habing, and $n_\mathrm{H} = \dens{4}$. The \R{12} models reach a comparable maximum abundance, $X(\ch{HCO+})_\mathrm{peak} = 1.87 \times 10^{-6}$, under nearly identical conditions, except for a slightly higher ionization rate of $\zeta = \crir[2.06]{-15}$ (Fig. \ref{fig:dense_gas_heatmap}).

Across the full parameter space, the peak abundances predicted by \R{22} agree with those from \R{12} to within 1~dex. In the majority of cases, $X(\ch{HCO+})$ lies between $10^{-7}$ and $10^{-6}$, with only the previously discussed global maxima exceeding this range. Although the peak conditions are similar, systematic differences between the two networks emerge at lower abundances, reflecting differences in the underlying chemical pathways.

The number of cases exhibiting significant differences, defined as $\Delta_\mathrm{R22/R12} \geq 1$~dex, increases with density (see Sect.~\ref{sec:diff-dense}). In several of these models, including the one with the highest $\Delta_\mathrm{R22/R12}$, the reaction
\begin{equation}
\label{re:carbon}
\ch{C + HCO+ -> CO + CH+},
\end{equation}
appears among the active destruction pathways in \R{12} at times when the \ch{HCO+} abundances begin to diverge between the two networks. This reaction is not included in \R{22}, suggesting that destruction by atomic carbon may contribute to the lower abundances predicted by \R{12}.

\subsection{Protostellar gas}
\label{sec:protostellar_gas}
\begin{figure*}[ht!]
    \centering
    \includegraphics[width=1\linewidth]{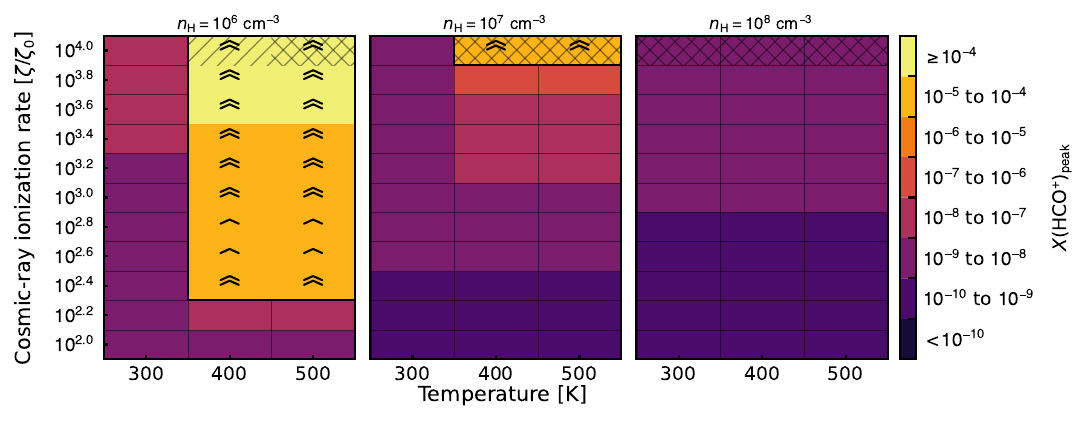}
    \caption{Same as Fig.~\ref{fig:dense_gas_heatmap}, but for protostellar gas. The x-axis shows the final temperature of the protostellar core, and each panel corresponds to a different density. In contrast to the dense molecular gas cases, models here reach $\Delta_\mathrm{R22/R12} \geq 1$~dex over extended regions of parameter space. A single chevron denotes $1 \le \Delta_\mathrm{R22/R12} < 2$~dex and a double chevron denotes $\Delta_\mathrm{R22/R12} \geq 2$~dex. Hatching marks regions where the peak \ch{HCO+} abundance is attained. These regions may occur at multiple temperatures within a given panel, indicating that the peak is tied to a specific temperature during the warm-up rather than to the final temperature.}
    \label{fig:protostellar_gas_heatmap}
\end{figure*}

In protostellar environments the gas undergoes a warm-up during which the temperature rises to several hundred Kelvin. Under these conditions the abundance of \ch{HCO+} becomes strongly temperature-dependent and can reach values several orders of magnitude higher than in the dense molecular gas models. 

The protostellar gas models considered in this work span temperatures up to 500~K and densities between $10^6$ and $\dens{8}$. Within this parameter space, extreme \ch{HCO+} enhancements occur only within a restricted combination of density, temperature, and CRIR, similar to those reported by D25, as seen in Fig.~\ref{fig:protostellar_gas_heatmap}. In particular, abundances exceeding $X(\ch{HCO+}) \geq 10^{-4}$ are reached only in a small subset of models with $n_\mathrm{H}=\dens{6}$ and $T_\mathrm{max}\geq 400$, where $\zeta/\zeta_0 \geq 10^{3.6}$. These extreme abundances are typically short-lived. An exception occurs for $T_{\rm max} = 400$~K with $\zeta/\zeta_0 \geq 10^{3.8}$, where once reached at $t\sim2\times10^5$~yr, they are sustained until the end of the model run at $10^6$~yr (Fig. \ref{fig:PG_400K}). 

We find that the density strongly limits the attainable \ch{HCO+} abundances. At $n_\mathrm{H} = \dens{8}$, the peak abundance, $X(\ch{HCO+})_\mathrm{peak}$, never exceeds $10^{-8}$. Moreover, across the densities, strong enhancements occur only in models with final temperatures between 400 and 500~K and $\zeta/\zeta_0 \geq 10^{2.4}$. The global maximum abundance in the \R{22} models is reached during the warm-up phase at $T \sim 440$~K with $X(\ch{HCO+})_\mathrm{peak} = 1.57\times10^{-4}$. This maximum occurs for the highest CRIR and lowest density. In contrast, the highest peak in \R{12} is more than 3~dex lower, with $X(\ch{HCO+})_\mathrm{peak} = 8.38\times10^{-7}$, and occurs at $T \sim 350$~K. This peak abundance is reached in models with final temperatures of both 400 and 500~K, indicating that it is set by a specific temperature during the warm-up rather than by the final temperature.

In the \R{22} model that produces the highest \ch{HCO+}, the maximum difference between the two networks reaches $\Delta_\mathrm{R22/R12}=3.98$~dex. An even larger difference $\Delta_\mathrm{R22/R12}=4.11$~dex occurs in the same setup with a slightly lower cosmic-ray ionization rate $\zeta/\zeta_0 = 10^{3.8}$. These large discrepancies arise at late evolutionary times when \ch{HCO+} becomes extremely abundant (Fig.~\ref{fig:PG_500K}). 

The analysis of the dominant reactions shows that at these timescales both networks share the same two dominant \ch{HCO+} destruction pathways:
\begin{equation}
\label{re:electron}
\ch{HCO+ + e- -> CO + H}
\end{equation}
\begin{equation}
\label{re:water}
\ch{H2O + HCO+ -> CO + H3O+}.
\end{equation}
Then the chemistry starts to diverge. In \R{22} the next important channel is the reaction of \ch{HCO+} with Mg, forming HCO and \ch{Mg+}. In \R{12}, an additional destruction pathway involving atomic carbon is present (Reaction~\ref{re:carbon}). The dominant formation pathways remain largely the same between the two networks. This suggests that the additional destruction channel in \R{12} likely contributes to the substantially lower \ch{HCO+} abundances predicted by that network, consistent with the behavior identified in the dense molecular gas models.

\begin{figure*}[ht!]
    \centering
    \includegraphics[width=.95\linewidth]{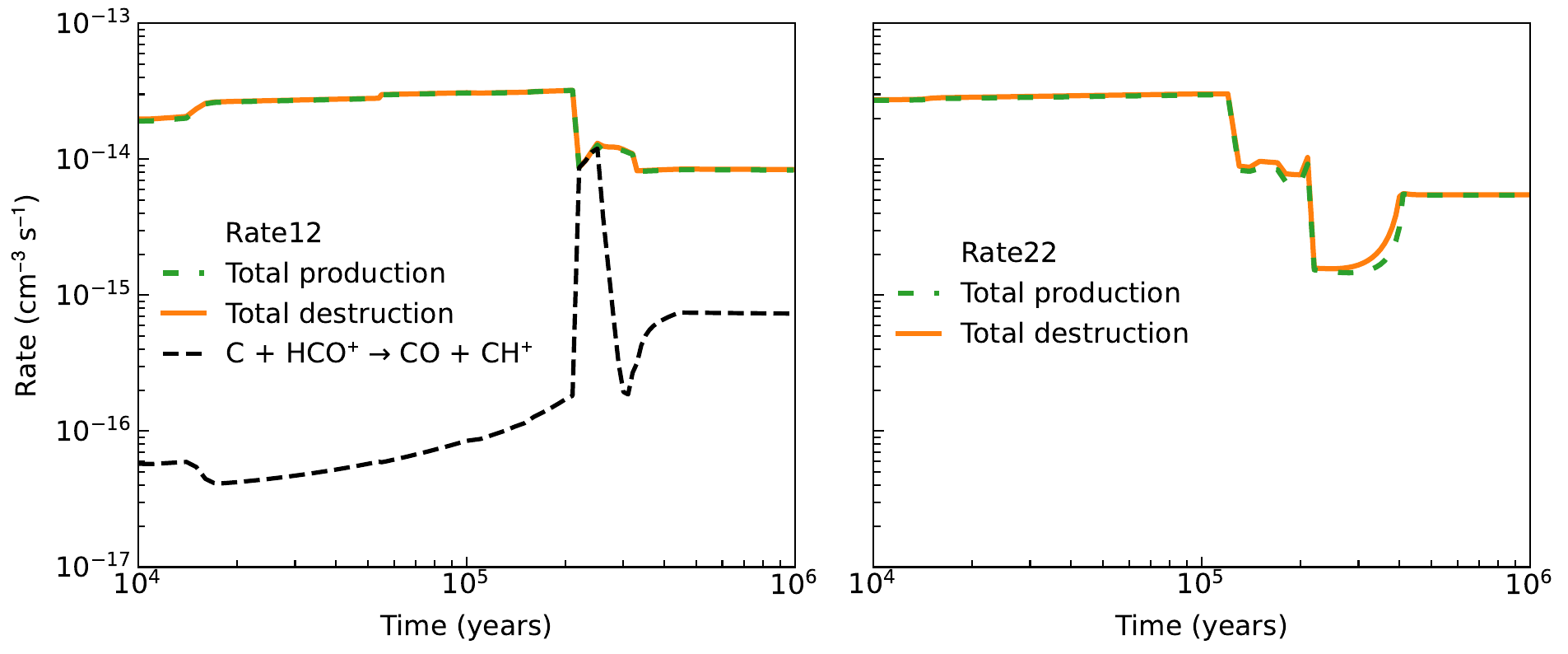}
    \caption{Total production and destruction rates in the protostellar gas model with the highest \ch{HCO+} abundance in \R{22} models ($n_{\rm H}=\dens{6}$, $T_{\rm final}=500$~K, and $\zeta=\crir{-13}$). The total rates are calculated from the five reactions that dominate the \ch{HCO+} chemistry at late times, when the extreme enhancement occurs. The reaction \ch{C + HCO+} is also shown where available.}
    \label{fig:reactions_protostellar}
\end{figure*}

For the model with the highest \ch{HCO+} abundance, we show the breakdown of the total production and destruction rates in \ch{HCO+} chemistry for the five dominant reactions in each channel, together with the rate of Reaction~\ref{re:carbon} (Fig.~\ref{fig:reactions_protostellar}). In both networks, production and destruction track each other closely throughout the evolution. The key difference emerges at $t \sim 2\times10^5$~yr, where the \ch{HCO+} abundance rises sharply in \R{22} (Fig.~\ref{fig:PG_500K}). In \R{12}, Reaction~\ref{re:carbon} spikes at precisely this moment, providing a destruction channel that suppresses the abundance enhancement. In \R{22}, where this reaction is absent\footnote{We note that this reaction is still included in the latest release of the KIDA chemical network \citep{wakelam2024}.}, no comparable channel operates and \ch{HCO+} reaches extreme abundances.

\subsection{Shocked gas}
\label{sect:fastshocks}

  \begin{figure*}[htbp]
      \centering
      \includegraphics[width=\textwidth]{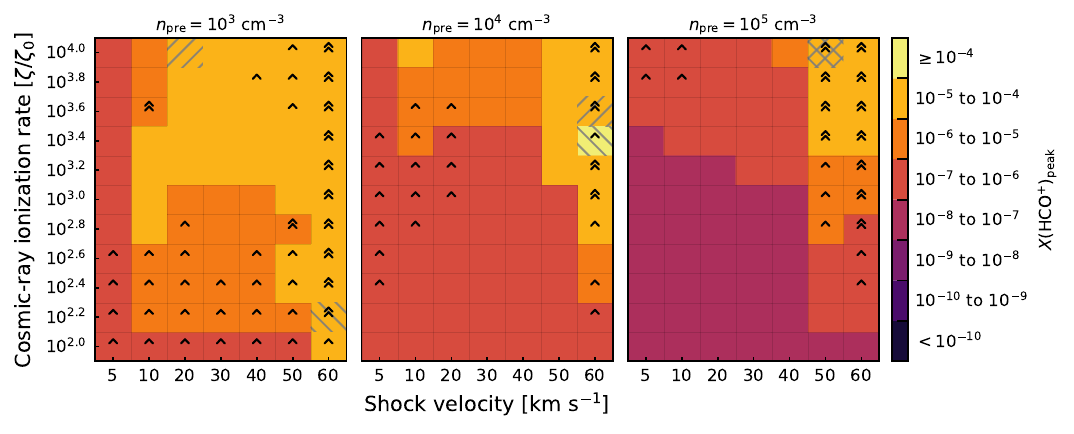}
      \caption{
      Same as Fig. \ref{fig:protostellar_gas_heatmap}, but for shocked gas. Each panel shows a different pre-shock density, with the x-axis corresponding to the shock velocity.
      }
      \label{fig:hcop-heatmap-umist22-vs-umist12}
    \end{figure*}

Figure~\ref{fig:hcop-heatmap-umist22-vs-umist12} presents the maximum fractional abundance of \ch{HCO+} reached in each model, projected onto the $(v_{\rm s},\,\zeta)$ plane for three values of the pre-shock density, $n_{\rm pre}=10^{3}$, $10^{4}$, and $10^{5}\,\mathrm{cm^{-3}}$.

At low pre-shock density, $n_{\rm pre}=10^{3}\,\mathrm{cm^{-3}}$, elevated peak abundances occupy a large fraction of the explored parameter space. For most values of $v_{\rm s}$, increasing $\zeta$ raises the maximum abundance from $\sim 10^{-6}$ to the $10^{-5}$--$10^{-4}$ range. Only models with low $\zeta$ and $\kms{5}$ remain below roughly $10^{-7}$--$10^{-6}$. In this regime, both networks predict broadly enhanced \ch{HCO+}, with peak abundances reaching $9.9\times10^{-5}$ in \R{22} and $6.1\times10^{-5}$ in \R{12}. The difference between the two networks becomes most pronounced toward high $v_{\rm s}$ and high $\zeta$, where it reaches $\Delta_{\mathrm{R22/R12}}=3.53$~dex.

\begin{figure*}[ht!]
    \centering
    \includegraphics[width=\linewidth]{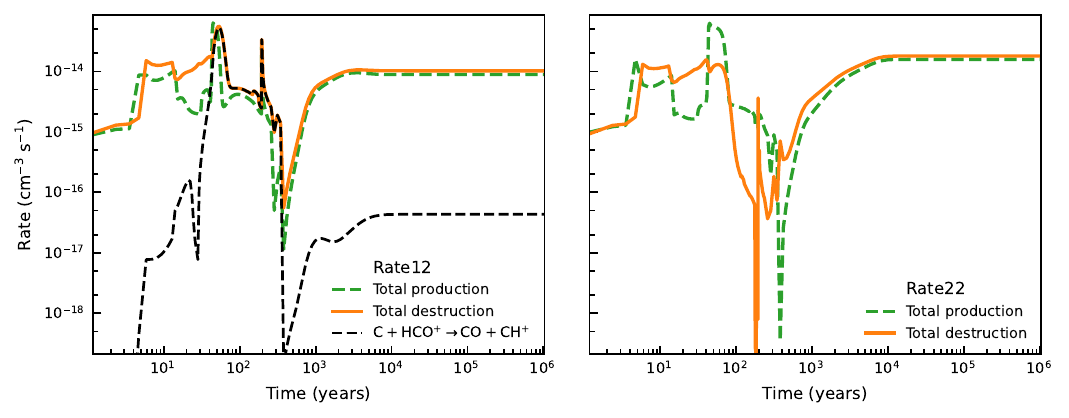}
    \caption{Same as Fig.~\ref{fig:reactions_protostellar}, but for shocked gas model ($n_{\rm pre}=\dens{4}$, $v_{\rm s}=\kms{60}$, and $\zeta=\crir[3.27]{-14}$), adjusted for the abundance enhancement timescales. Differences between the total formation and destruction rates reflect the non-equilibrium nature of the shock chemistry.}
    \label{fig:reactions_shocks}
\end{figure*}

For $n_{\rm pre}=10^{4}\,\mathrm{cm^{-3}}$, the region producing high $X(\ch{HCO+})_{\rm peak}$ becomes noticeably reduced compared to the $10^{3}\,\mathrm{cm^{-3}}$ case. Abundances above $10^{-6}$ are typically found at elevated $\zeta$ and moderate to high $v_{\rm s}$, whereas lower $\zeta$ models remain at $\lesssim 10^{-7}$. The maximum abundance in this panel is reached in \R{22}, with $X(\ch{HCO+})_{\rm peak}=1.0\times10^{-4}$, which also corresponds to the highest value found in the full set of shock models, while the corresponding maximum in \R{12} is $5.0\times10^{-5}$. As in the lower density case, the strongest difference between the two networks appears in the fast, strongly ionized shocks, where $\Delta_{\mathrm{R22/R12}}=2.99$~dex.

At the highest pre-shock density, $n_{\rm pre}=10^{5}\,\mathrm{cm^{-3}}$, the parameter space producing very high peak abundances becomes much more restricted. Only models with both high $v_{\rm s}$ and high $\zeta$ approach or exceed $10^{-6}$, while a substantial fraction of models remain below $10^{-8}$. This trend is consistent with the cloud and protostellar models presented above, where high gas density also suppresses $X(\ch{HCO+})_{\rm peak}$. The maximum abundance predicted by \R{22} is $6.7\times10^{-5}$, compared to $1.2\times10^{-5}$ in \R{12}. The largest network difference is again found in the fast shock regime, where the abundance contrast reaches $\Delta_{\mathrm{R22/R12}}~=~3.73$~dex.

Consistent with the other models investigated in this work, we find that the removal of the destruction pathway \ref{re:carbon} appears to be the main factor leading to the extremely high \ch{HCO+} abundances in these cases. Figure~\ref{fig:reactions_shocks} presents the production and destruction reaction rates of \ch{HCO+} for \R{12} and \R{22}, shown in the left and right columns, respectively. The model corresponds to the parameter combination that produces the highest \ch{HCO+} abundance in \R{22} in Fig.~\ref{fig:hcop-heatmap-umist22-vs-umist12}. 

The left panel shows that the destruction rate of \ch{HCO+} exhibits two prominent peaks around 80 and 300 years, during which it dominates the overall destruction budget of \ch{HCO+}. These intervals coincide with the period when the abundance difference between \R{12} and \R{22} becomes most pronounced. In contrast, neither of these destruction peaks appears in the right panel for \R{22}, indicating that no comparable destruction channel operates during the same phases. This comparison suggests that the extremely high $X(\ch{HCO+})$ values in \R{22} arise primarily from the absence of this efficient destruction pathway, allowing \ch{HCO+} to survive longer and accumulate to much higher abundances.

\subsection{Electron fraction}

\begin{figure}[ht!]
    \centering
    \includegraphics[width=1\linewidth]{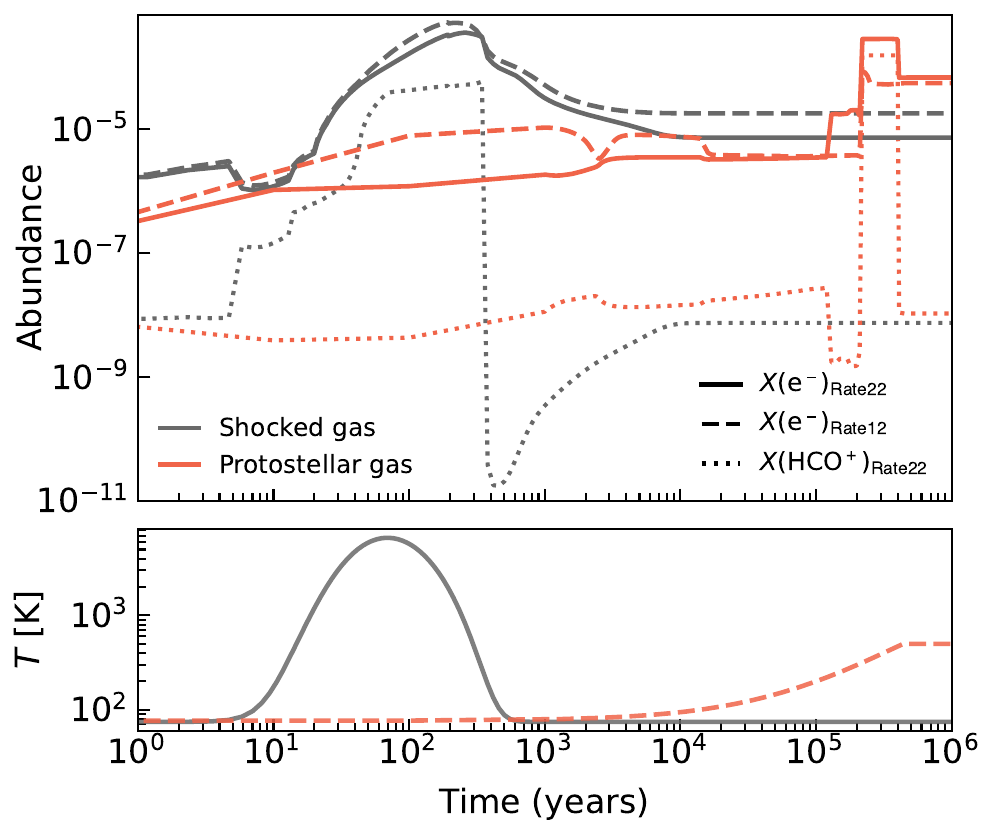}
    \caption{Electron fraction as a function of time in the shocked (grey; $n_{\rm pre}=\dens{4}$, $v_{\rm s}= \kms{60}$, and $\zeta=\crir[3.27]{-14}$) and protostellar (orange; $n_{\rm H}=\dens{6}$, $T_{\rm final}=500$~K, and $\zeta=\crir{-13}$) gas models exhibiting the highest \ch{HCO+} enhancement. Dotted lines show the corresponding \ch{HCO+} abundances predicted by the \R{22} network. The bottom panels show the corresponding temperature profiles. Differences in the electron fraction are present between the two chemical networks, but are substantially smaller than those found for \ch{HCO+}.}
    \label{fig:electrons_temperature}
\end{figure}

For the models with the highest \ch{HCO+} abundance predicted by the \R{22} network in shocked and protostellar gas, we examined the electron fraction to determine whether significant changes coincide with the periods of enhanced \ch{HCO+} abundance (Fig.~\ref{fig:electrons_temperature}). This is particularly relevant because \ch{HCO+} is efficiently destroyed through dissociative recombination with electrons (Reaction \ref{re:electron}), which is among the dominant destruction pathways in both chemical networks (Sect.~\ref{sec:protostellar_gas}).

We find that the electron fraction differs between the two networks throughout the evolution, including during the periods of enhanced \ch{HCO+} abundance in both the shocked and protostellar models. These differences are considerably smaller than those found for \ch{HCO+} and many other chemically related species, as discussed in Sect.~\ref{sec:other_molecules}. Nevertheless, this demonstrates that the choice of the chemical network also affects the overall ionization balance. Such differences could be particularly relevant in models, where ionization fraction is linked to the physical evolution through the magnetic field.

\subsection{Chemical uncertainties and observational considerations}

Some of the abundances presented in this work, particularly related to the \R{22} network, reach extremely high values of $X(\ch{HCO+})\gtrsim 10^{-4}$. Such abundances cannot represent the bulk conditions of molecular gas, but instead could correspond to short-lived or localized regions where the combination of high temperature, density, and ionization rate strongly enhances the \ch{HCO+} chemistry. Environments with these properties are more likely to occur in extreme regions like starburst nuclei or active galactic nuclei-affected gas than in typical Galactic molecular clouds.

In practice, observed emission from \ch{HCO+} traces gas integrated along the line of sight and within the telescope beam. The inferred abundances therefore represent an average over multiple gas components with different physical conditions. Localized regions with very high \ch{HCO+} abundance would likely be embedded within larger volumes of gas with lower abundances, resulting in beam-averaged values closer to the commonly observed range of $10^{-9}-10^{-8}$. This raises the question of whether such extreme abundances occur in astrophysical environments but remain observationally unresolved, or whether they instead reflect gaps in the current understanding of the chemistry of \ch{HCO+}.

As discussed in previous sections, the main chemical difference appears to be related to the treatment of an important destruction pathway of \ch{HCO+}, namely the reaction \ch{C + HCO+ -> CO + CH+}, which is present in \R{12} but absent in \R{22}. This interpretation is supported by the fact that reinstating this reaction in the \R{22} network suppresses the extreme \ch{HCO+} enhancement. 

The inclusion of this reaction in astrochemical databases has been the subject of ongoing debate. It was originally assumed to be exoergic \citep{prasad1980}. However, more recent theoretical studies suggest that the direct product channels are strongly endoergic (by at least 8402~K) and associated with substantial activation barriers \citep{savin2017}. If only these direct reaction channels are considered, the reaction would be expected to be strongly suppressed under typical dark cloud conditions.

The UMIST database is, however, widely applied to model a broad range of astrophysical environments, often very different from dark clouds \citep[e.g.,][]{booth2021,luo2024,vandesande2024}. If a substantial activation barrier is present, the reaction may be suppressed in cold gas, but it could still proceed in higher-temperature environments where the barrier can be overcome. For example, fast shocks can reach temperatures exceeding $10^4$~K \citep[e.g.,][]{godard2019}, under such conditions this reaction may contribute to the destruction of \ch{HCO+}. 

It is also possible that the reaction mechanism is more complex than a simple barrier-mediated process. One possible source of this uncertainty is intersystem crossing between different electronic spin surfaces, which may alter the reaction pathways available to the system. Consequently, the effective low-temperature rate coefficient remains uncertain. Further theoretical and experimental studies are therefore needed to better constrain the reaction mechanism and the corresponding rate coefficient.

If, on the other hand, the chemistry adopted in \R{22} provides a more accurate description of \ch{HCO+} formation and destruction, the focus shifts from chemistry to observational tests. This raises the question of why such enhanced \ch{HCO+} abundances have not yet been reported in environments where the required physical conditions are expected to occur. The extreme abundances would most likely be confined to highly localized environments associated with shocks or protostellar activity, but also regions such as the circumnuclear disk of Sgr A$^*$, where localized pockets of gas with physical conditions comparable to those producing enhanced \ch{HCO+} in our models have been inferred \citep{james2021}. Since such regions would be difficult to spatially resolve in extragalactic observations, their signatures would likely be diluted by emission from the surrounding gas with lower \ch{HCO+} abundances, making them challenging to detect. The most promising locations to search for these extreme conditions may be within our own Galaxy, where higher spatial resolution can be achieved. Regions such as the Galactic Center, which host dense gas exposed to strong heating and ionization, may therefore provide the best opportunity to test the chemical predictions of the \R{22} network.

\subsection{Impact on other molecules}
\label{sec:other_molecules}

\begin{figure*}[ht!]
    \centering
    \includegraphics[width=.49\linewidth]{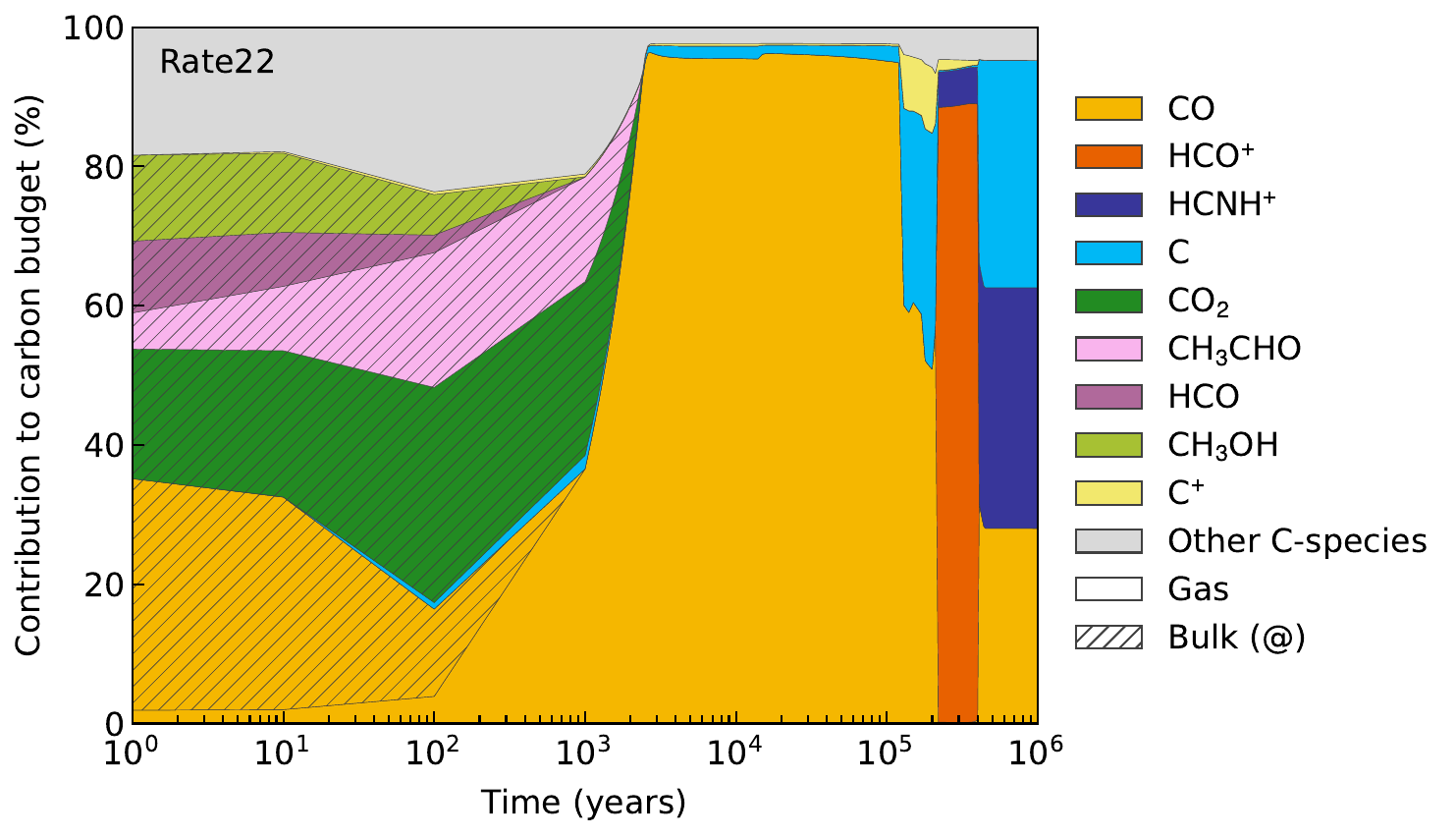}
    \includegraphics[width=0.49\linewidth]{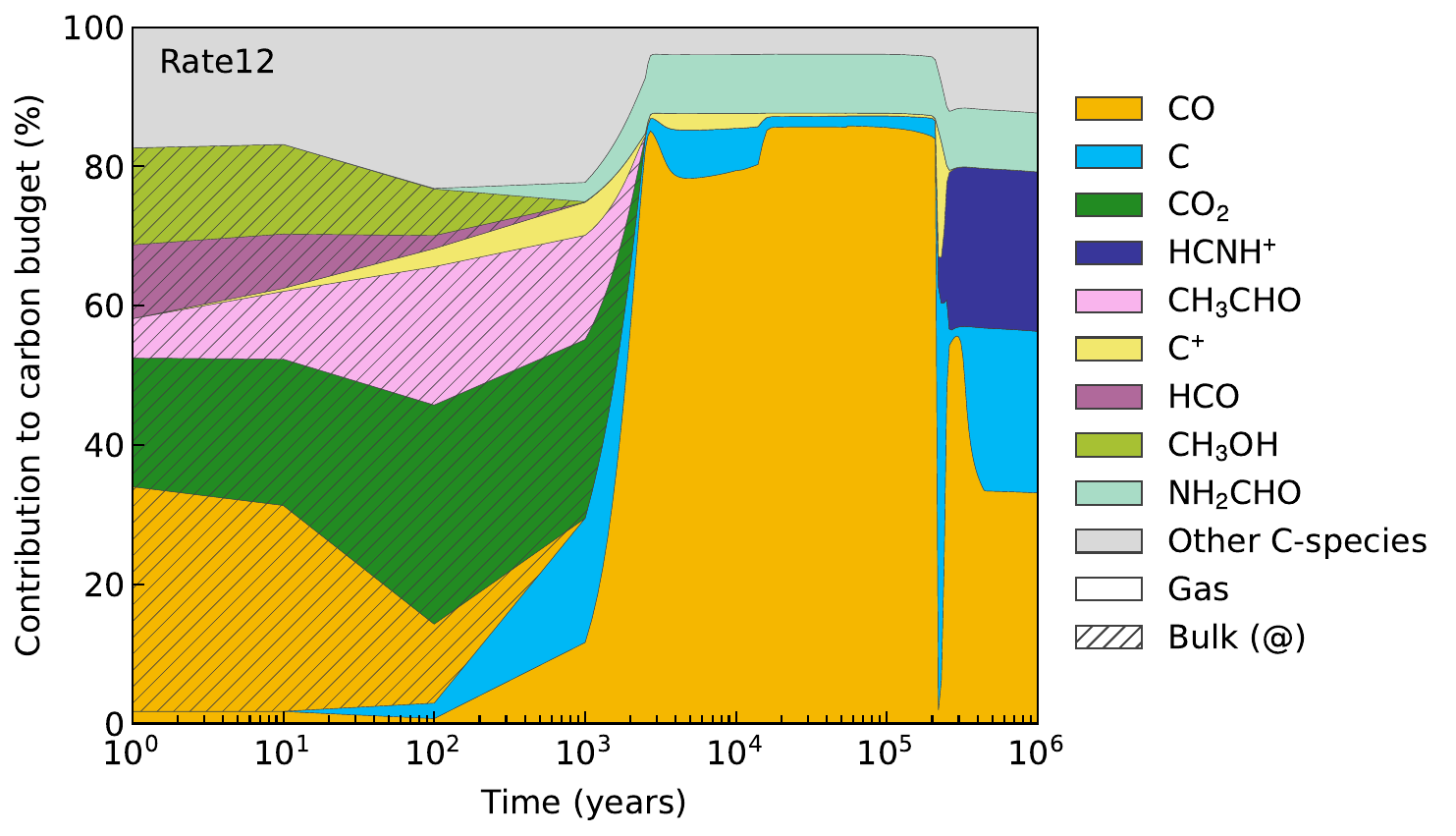}
    \caption{Fractional distribution of the carbon budget in the protostellar gas model that produces the highest \ch{HCO+} abundance within the explored parameter space ($n_{\rm H}=\dens{6}$, $T_{\rm final}=500$~K, and $\zeta=\crir{-13}$). The left panel shows the \R{22} results and the right panel shows the corresponding \R{12} results. The vertical thickness of each band represents the fraction of the total carbon budget carried by a given species, calculated as $N_\mathrm{C}(i)X_i /\sum_j N_\mathrm{C}(j)X_j$, where $X_i$ is the abundance of species $i$ relative to hydrogen nuclei and $N_\mathrm{C}(i)$ is the number of carbon atoms in that species. Hatched regions indicate bulk-ice species residing in the mantle beneath the grain surface, while unhatched regions correspond to gas-phase species.}
    \label{fig:carbon_budget}
\end{figure*}

Although a detailed analysis of other species is beyond the scope of this work, we briefly examined whether the extreme \ch{HCO+} enhancement identified in our models affects the abundances of other molecules. The most direct impact is expected for \ch{CO}, which is typically the dominant reservoir of carbon in molecular gas. In the models where $X({\ch{HCO+}}) \gtrsim  10^{-4}$, the abundance of \ch{HCO+} becomes comparable to that of initial carbon reservoir, implying that a significant fraction of carbon may temporarily reside in \ch{HCO+} rather than \ch{CO}. Under such conditions, the abundances of other chemically related species may also be affected.

To investigate the carbon balance, we examined the model producing the highest \ch{HCO+} abundance ($n_{\rm H}=\dens{6}$, $T_{\rm final}=500$~K, and $\zeta=\crir{-13}$), shown in Fig.~\ref{fig:carbon_budget}. While CO is the dominant carbon reservoir, the \ch{HCO+} enhancement in \R{22} is accompanied by a significant redistribution of the carbon budget, with \ch{HCO+} temporarily becoming the dominant carbon-bearing species. In contrast, \R{12} does not exhibit such a redistribution, and \ch{HCO+} remains only a minor contributor to the overall carbon budget. Instead, carbon is retained primarily in CO and atomic C, which is consistent with the presence of Reaction~\ref{re:carbon} in \R{12}, efficiently converting \ch{HCO+} back into CO.

The redistribution of the carbon budget during the \ch{HCO+} enhancement is closely linked to changes in the abundances of other chemically related species. We therefore examined the behavior of commonly observed molecules associated with this chemistry in the model that produces the highest \ch{HCO+} abundance. In particular, we focused on molecules that participate in the dominant formation and destruction pathways of \ch{HCO+}, including \ch{CO}, \ch{CH+}, \ch{OH}, \ch{OH+}, \ch{H3+}, \ch{N2H+}, \ch{HCO}, \ch{HOC+}, \ch{H2O}, and \ch{H3O+}.

\begin{figure*}[ht!]
    \centering
    \includegraphics[width=1\linewidth]{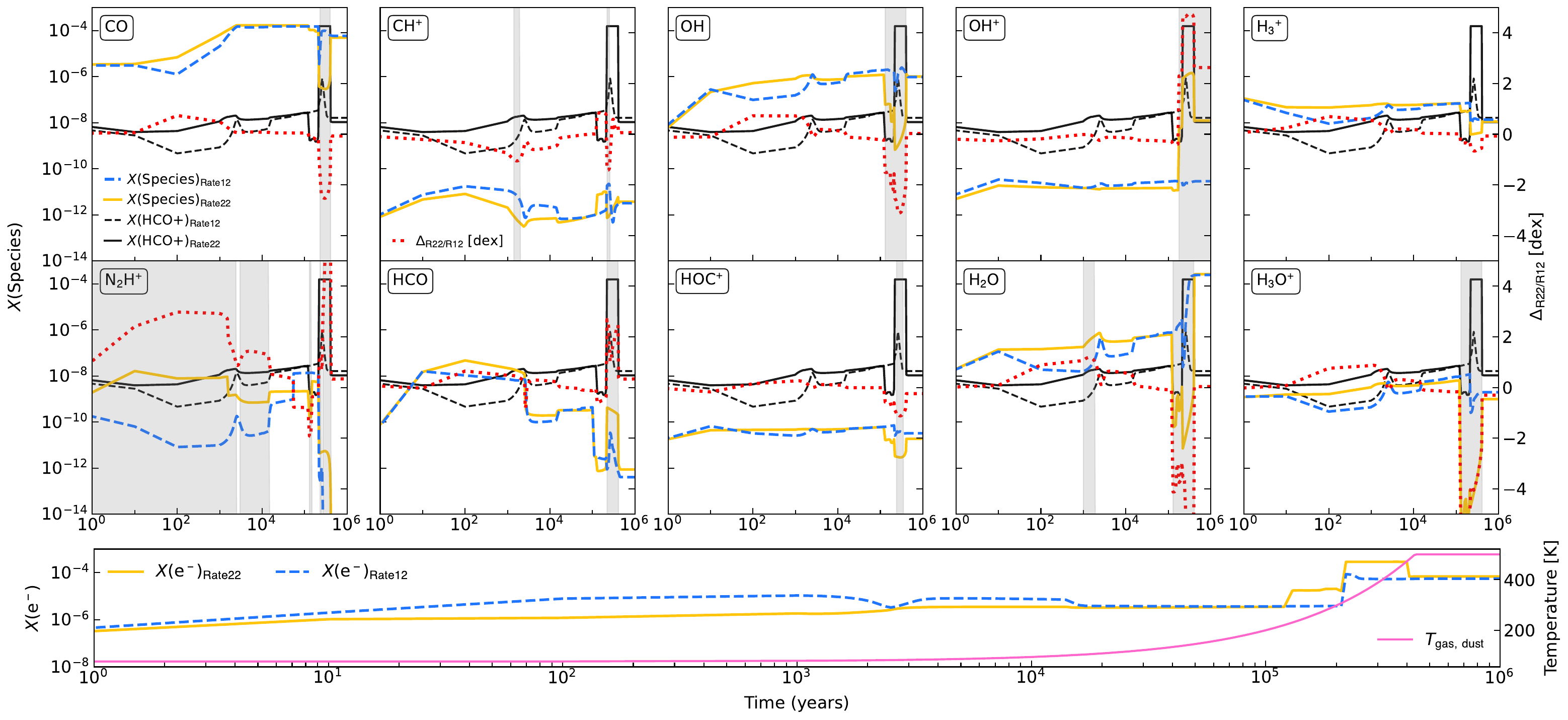}
    \caption{Chemical evolution of molecules associated with the dominant \ch{HCO+} chemistry for the same model as shown in Fig.~\ref{fig:carbon_budget}. The figure shows the chemical evolution in the protostellar gas model that produces the highest \ch{HCO+} abundance within the explored parameter space. Solid and dashed lines correspond to the \R{22} and \R{12} chemical networks, respectively. In each panel, the abundance of \ch{HCO+} is overplotted to indicate the time of extreme enhancement. The dotted red line shows the abundance difference $\Delta_\mathrm{R22/R12}$, while grey shading marks periods where the difference exceeds 1~dex. The bottom panel shows the evolution of the electron fraction and gas temperature. Most species exhibit differences larger than 1~dex between the two networks, with the exception of \ch{H3+}. The largest deviations occur near the time of peak \ch{HCO+} abundance, with particularly strong variations for \ch{N2H+}, \ch{H2O}, and \ch{H3O+}, where the differences exceed 4~dex.}
    \label{fig:other_molecules}
\end{figure*}

Most of these species exhibit abundance differences exceeding 1~dex between the two chemical networks, with the exception of \ch{H3+} (Fig.~\ref{fig:other_molecules}). The largest deviations typically occur near the time when \ch{HCO+} undergoes its extreme enhancement. However, it is difficult to isolate whether these differences arise directly from the enhanced \ch{HCO+} chemistry or from other variations between \R{22} and \R{12}. Some species also show noticeable differences at earlier stages of the evolution, suggesting that additional chemical pathways may contribute to these discrepancies. We find that the most extreme differences exceeding 4~dex are observed for \ch{N2H+}, \ch{H2O}, and \ch{H3O+}. This indicates that the chemistry in this physical regime is strongly influenced by the treatment of key ion-neutral reactions.

\section{Conclusions}
\label{sect:conclusions}

This work revisits the extremely high \ch{HCO+} abundances reported by D25 in the high-$\zeta$ regime. We investigated the physical conditions under which such enhancements arise and assessed their robustness across different astrophysical environments. In addition to protostellar and shocked gas, we also examined dense molecular gas conditions. To evaluate the impact of gas-phase chemistry, all models were computed using two chemical networks: \R{12} \citep{umist2012} and \R{22} \citep{umist2022}.

We find that strong \ch{HCO+} enhancement, reaching $X(\ch{HCO+}) \gtrsim 10^{-4}$, is possible in protostellar and shocked gas under specific combinations of temperature, density, and cosmic-ray ionization rate. In general, increasing density suppresses the peak \ch{HCO+} abundance, requiring higher ionization rates to achieve comparable enhancements. Shocks tend to produce higher average abundances, while the most extreme peak values occur in the protostellar gas models.

Comparing the two chemical networks shows that the largest discrepancies arise from differences in the destruction pathways of \ch{HCO+}. In particular, the absence of the reaction C~+~\ch{HCO+}~$\longrightarrow$~CO~+~\ch{CH+} in \R{22} leads to substantially higher \ch{HCO+} abundances than in \R{12}. In the most extreme cases, the predicted abundances differ by more than 4~dex between the networks. These differences also propagate to other molecules chemically linked to \ch{HCO+}, indicating that the chemistry in this regime is highly sensitive to the ion-neutral reaction network.

Because \R{22} represents a more recent update of the UMIST database, these results highlight the importance of carefully comparing predictions obtained with different chemical networks when studying environments characterized by high temperatures and elevated cosmic-ray ionization rates. In such regimes, small differences in dominant reactions can lead to large changes in the predicted abundances of several commonly observed molecules. More generally, the large network differences found here raise the possibility that relevant reactions may be missing from current chemical databases under extreme physical conditions, particularly in regimes where networks have not been fully tested against observations. Future studies should therefore revisit the chemistry of other species under similar conditions and avoid interpreting individual molecular tracers in isolation.

\begin{acknowledgements}
We thank Tom Millar and Marie Van de Sande for helpful clarifications regarding the latest UMIST release, Jeff Mangum for insightful discussions on the potential observational implications of high \ch{HCO+} abundances, and François Lique for valuable insights into the destruction channels of \ch{HCO+}.

KMD and SV are funded by the European Research Council (ERC) Advanced Grant MOPPEX 833460. BJ acknowledges support from the CSC (China Scholarship Council) scholarship program. TMD is financially supported by the Dutch Astrochemistry Network of the Dutch Research Council (NWO) under grant no.~ASTRO.JWST.001. 
\end{acknowledgements}

\bibliographystyle{aa}
\bibliography{references}

\begin{appendix}

\section{Differences in abundances between networks}
\label{sec:differnces}

As part of this study, we examined how the choice of gas-phase chemical network affected the predicted \ch{HCO+} abundances. Previous work (D25) has shown that variations in the adopted chemistry can significantly influence \ch{HCO+} under specific physical conditions. To assess the impact of these changes, we computed the same grid of models using both networks, \R{12} and \R{22}. We quantified the difference between the predicted abundances as $\Delta_\mathrm{R22/R12}$, expressed in dex, and focused on cases where $\Delta_\mathrm{R22/R12} \geq 1$~dex, corresponding to at least an order-of-magnitude difference. We discuss these differences for dense molecular gas in Sect.~\ref{sec:diff-dense}, protostellar gas in Sect.~\ref{sec:diff-protostellar}, and shocked gas in Sect.~\ref{sec:diff-shocked}.

\subsection{Dense molecular gas}
\label{sec:diff-dense}
\begin{figure*}[ht]
    \centering
    \subfigure{\includegraphics[width=0.33\textwidth]{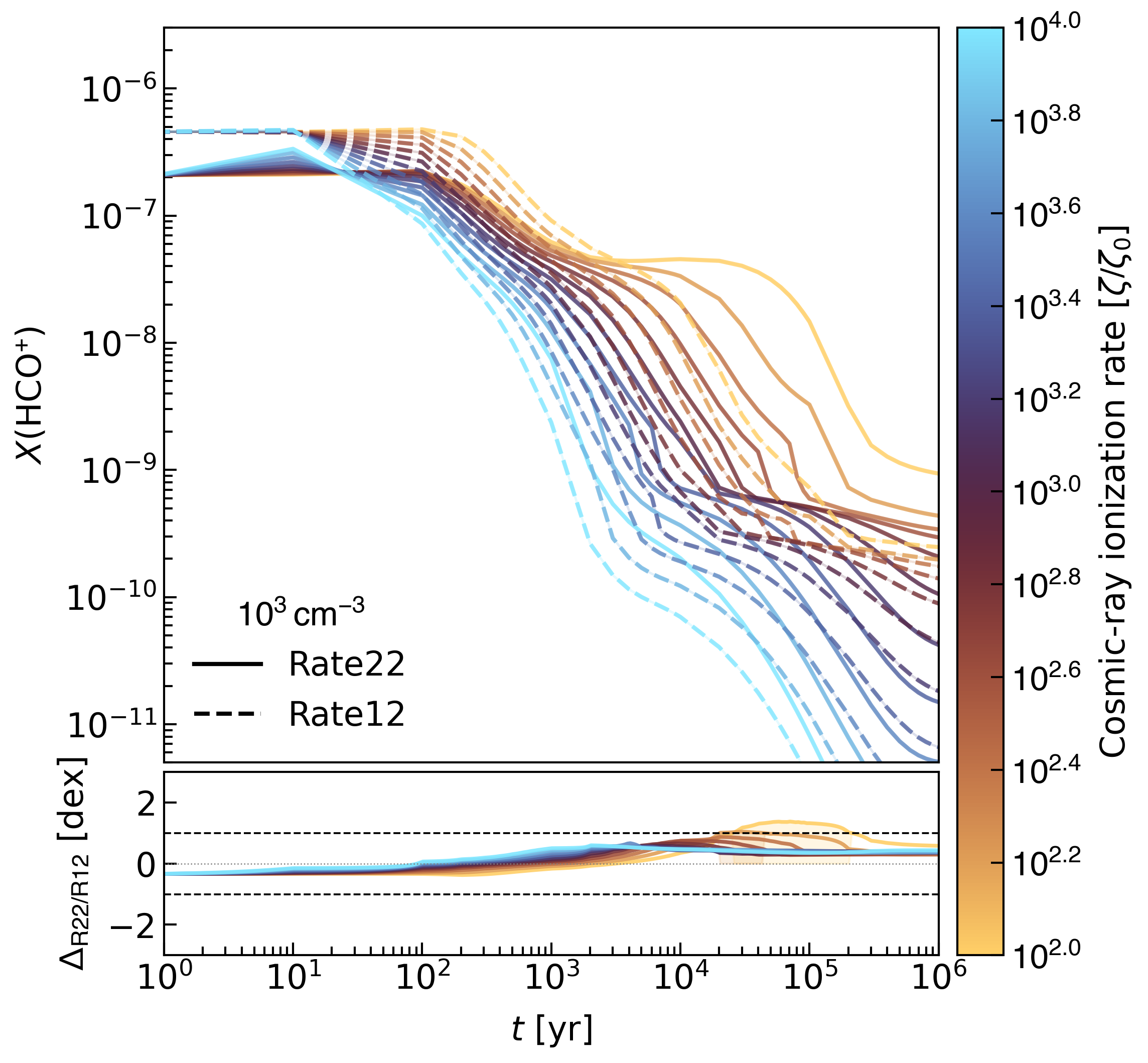}}
    \subfigure{\includegraphics[width=0.33\textwidth]{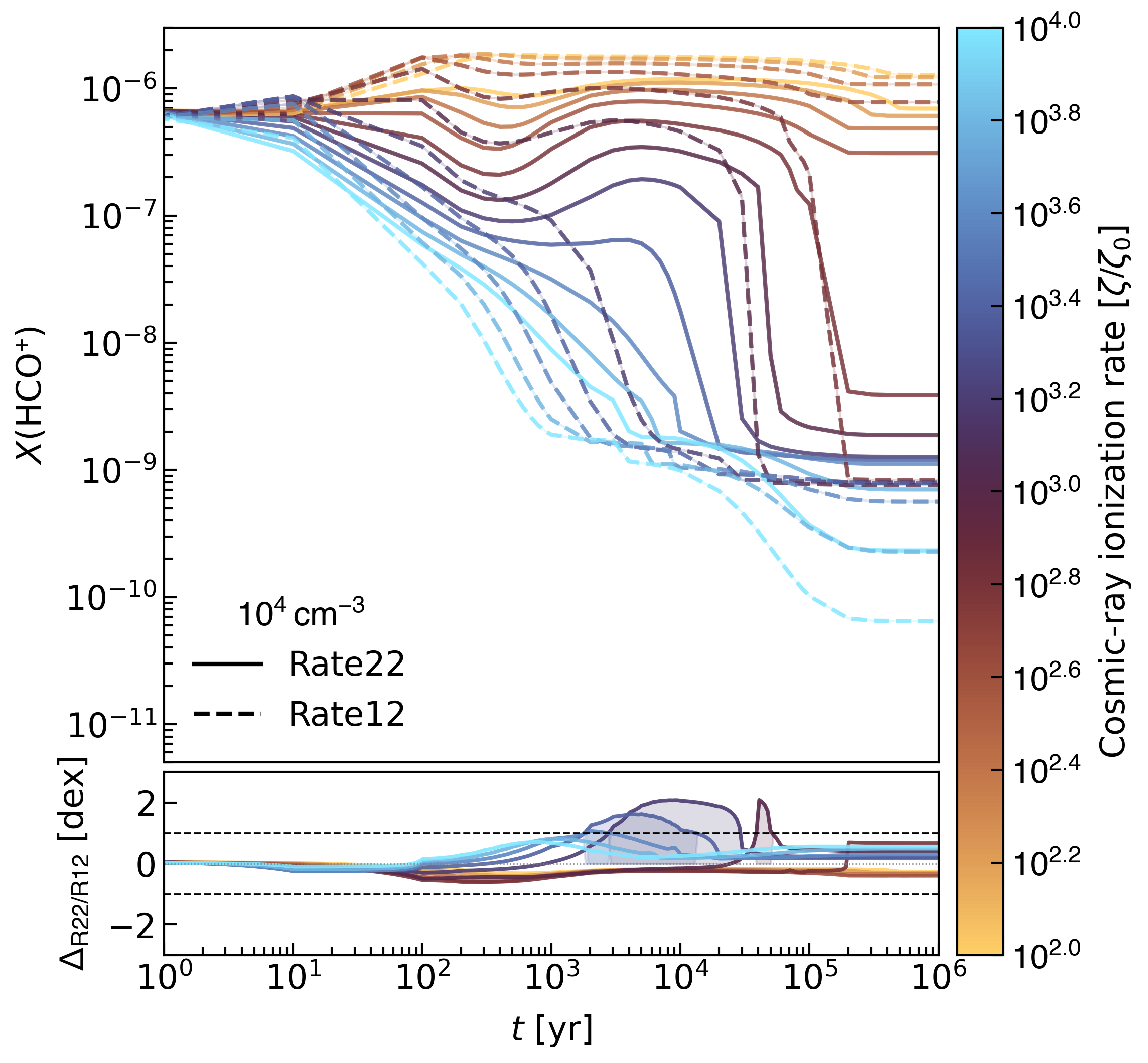}}
    \vfill
    \subfigure{\includegraphics[width=0.33\textwidth]{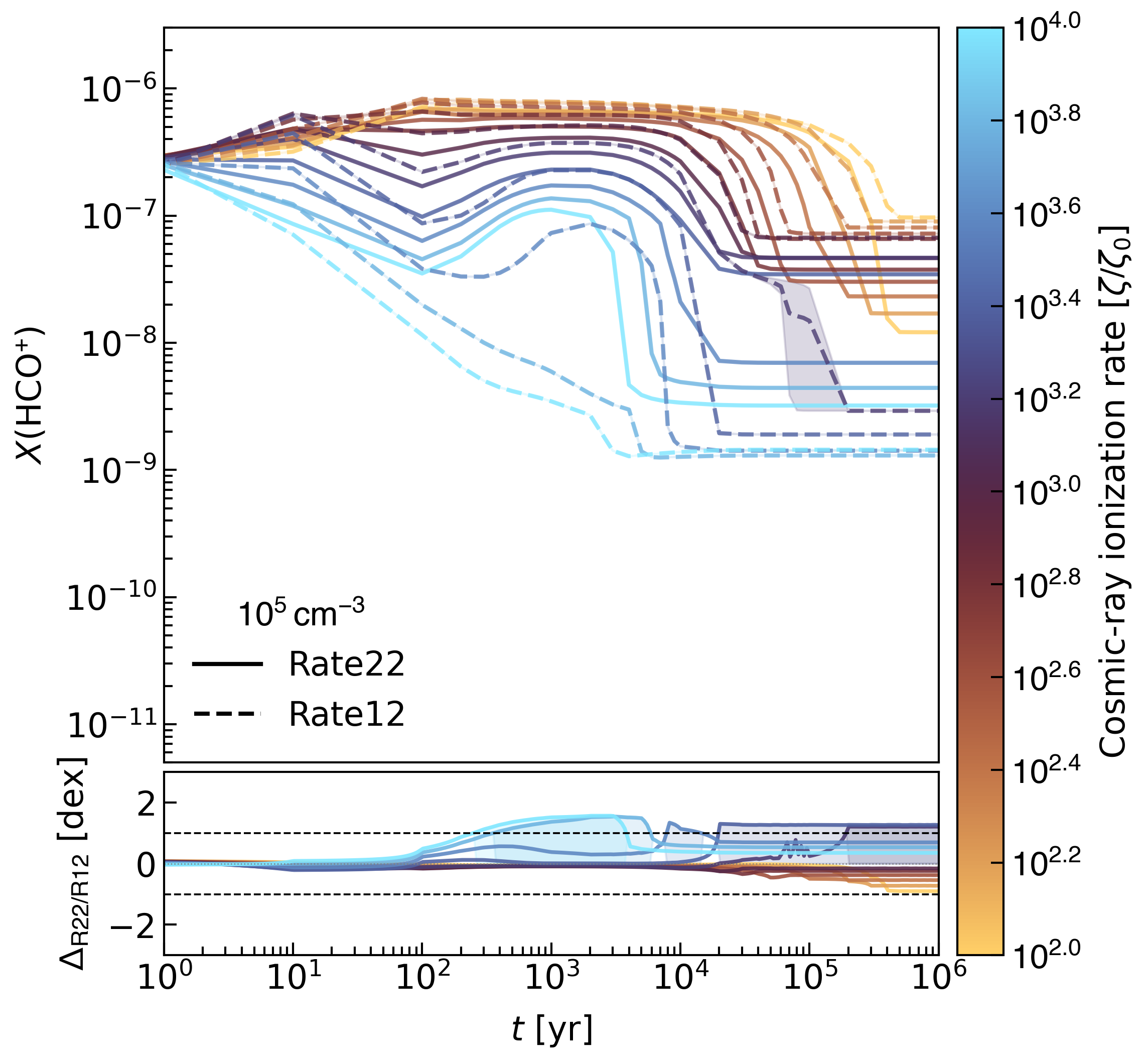}}
    \subfigure{\includegraphics[width=0.33\textwidth]{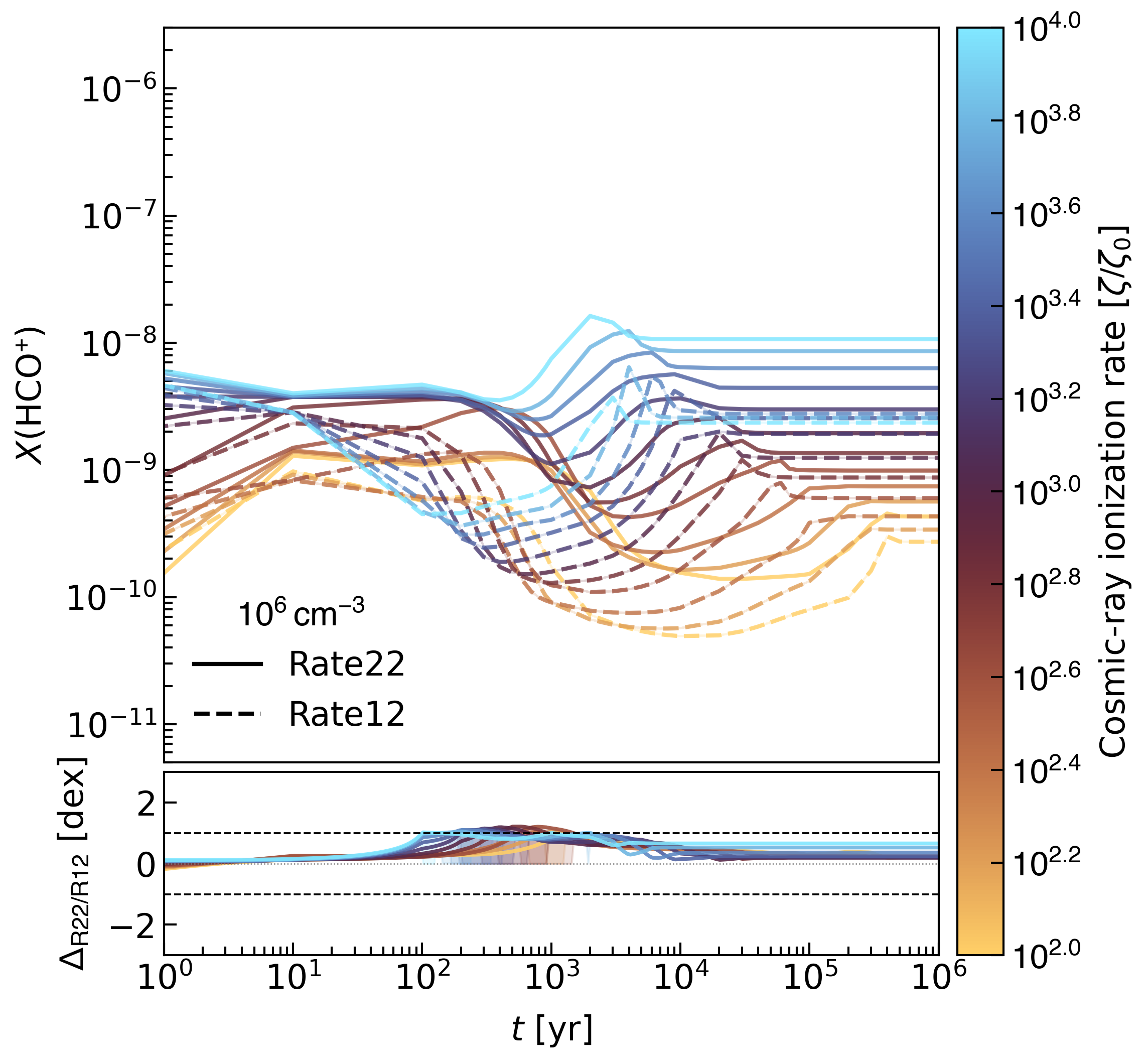}}
    \label{fig:DMG_all}
    \caption{Evolution of the \ch{HCO+} abundance in dense molecular gas models for $n_{\rm H}=10^3-\dens{6}$. The top panels show the fractional abundance $X(\ch{HCO+})$ predicted by \R{22} (solid lines) and \R{12} (dashed lines) for a range of cosmic-ray ionization rates, $\zeta/\zeta_0$, indicated by different colors. The bottom panels show the corresponding abundance difference $\Delta_\mathrm{R22/R12}$ (in dex), with dotted horizontal lines marking differences of one order of magnitude (1~dex). The number of cases exhibiting significant differences between the networks increases with density, with the largest difference of 2.07~dex found for $n_{\rm H}=\dens{4}$.}
\end{figure*}

Dense molecular gas models correspond to static clouds with temperatures set by the cosmic ray ionization rate, and span densities of $n_\mathrm{H} = 10^3-\dens{6}$. We find that the number of cases exhibiting significant differences between the two chemical networks increases with density (Fig. \ref{fig:DMG_all}). This increase does not directly translate into the magnitude of the differences. Instead, \R{22} tends to maintain higher abundances in denser gas under otherwise identical physical conditions. The largest difference, $\Delta_\mathrm{R22/R12}=2.07$~dex, is found for $n_\mathrm{H}=\dens{4}$ and $\zeta = \crir[2.06]{-14}$. At the same time, the $n_\mathrm{H}=\dens{5}$ models sustain $\Delta_\mathrm{R22/R12}\geq1$~dex for the longest duration, persisting up to $9.80\times10^5$~years. 

The network differences emerge after $\sim$100~yr, with both networks predicting broadly similar \ch{HCO+} abundances at earlier times. Beyond this point, \R{22} generally produces higher \ch{HCO+} abundances at mid-to-high $\zeta$, which is consistent with the removal of the \ch{C + HCO+ -> CO + CH+} destruction pathway. At low $\zeta$, the behavior is more complex and in some cases reversed, particularly at $n_\mathrm{H}=\dens{3}$ and at late times for $n_\mathrm{H}=\dens{5}$. This indicates that the impact of individual network differences depends on the local physical conditions, with density and ionization rate jointly determining which network differences are most consequential in the case of \ch{HCO+} chemistry.

\subsection{Protostellar gas}
\label{sec:diff-protostellar}
\begin{figure*}[ht]
    \centering
    \subfigure{\includegraphics[width=0.33\textwidth]{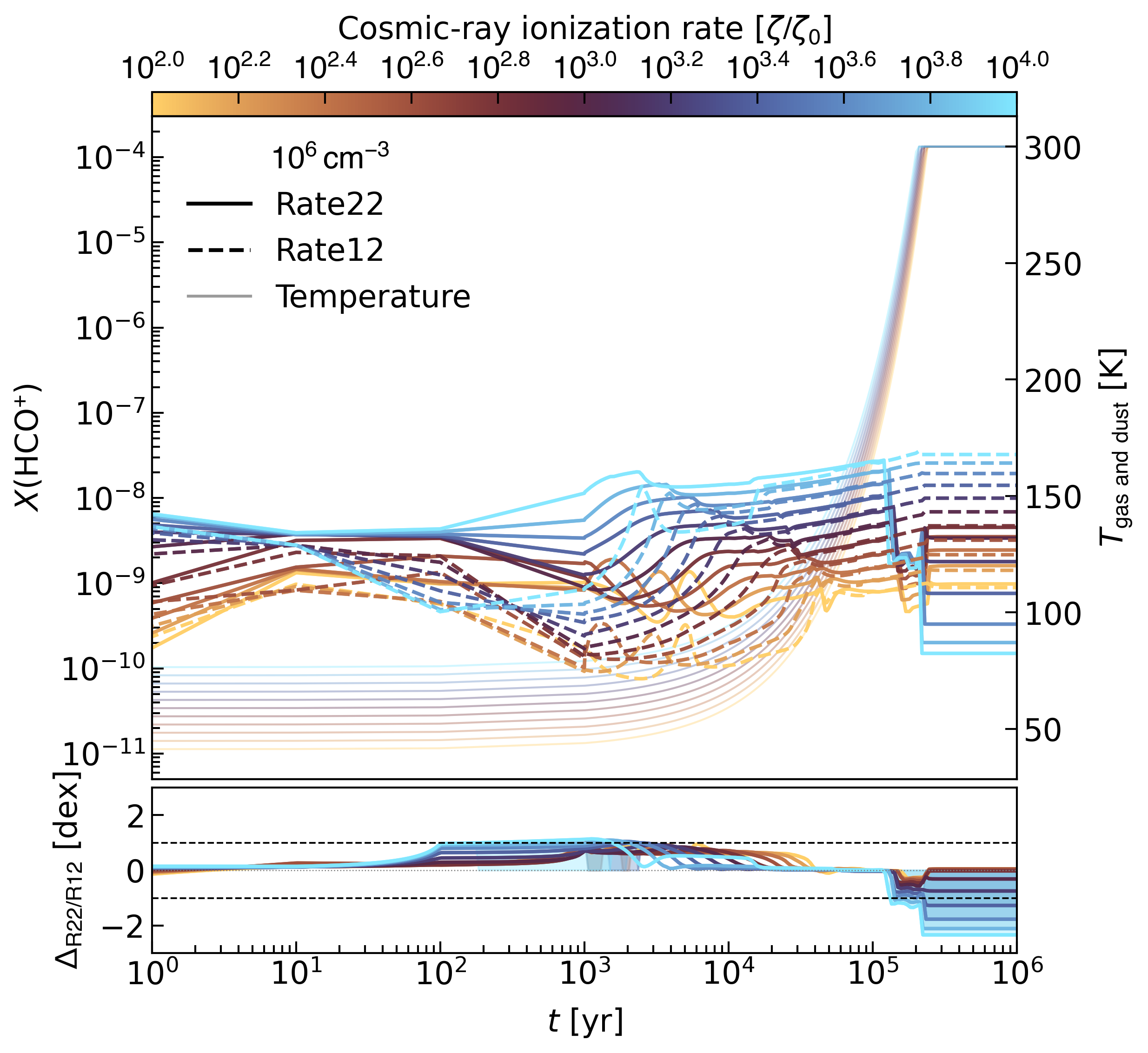}}
    \hfill
    \subfigure{\includegraphics[width=0.33\textwidth]{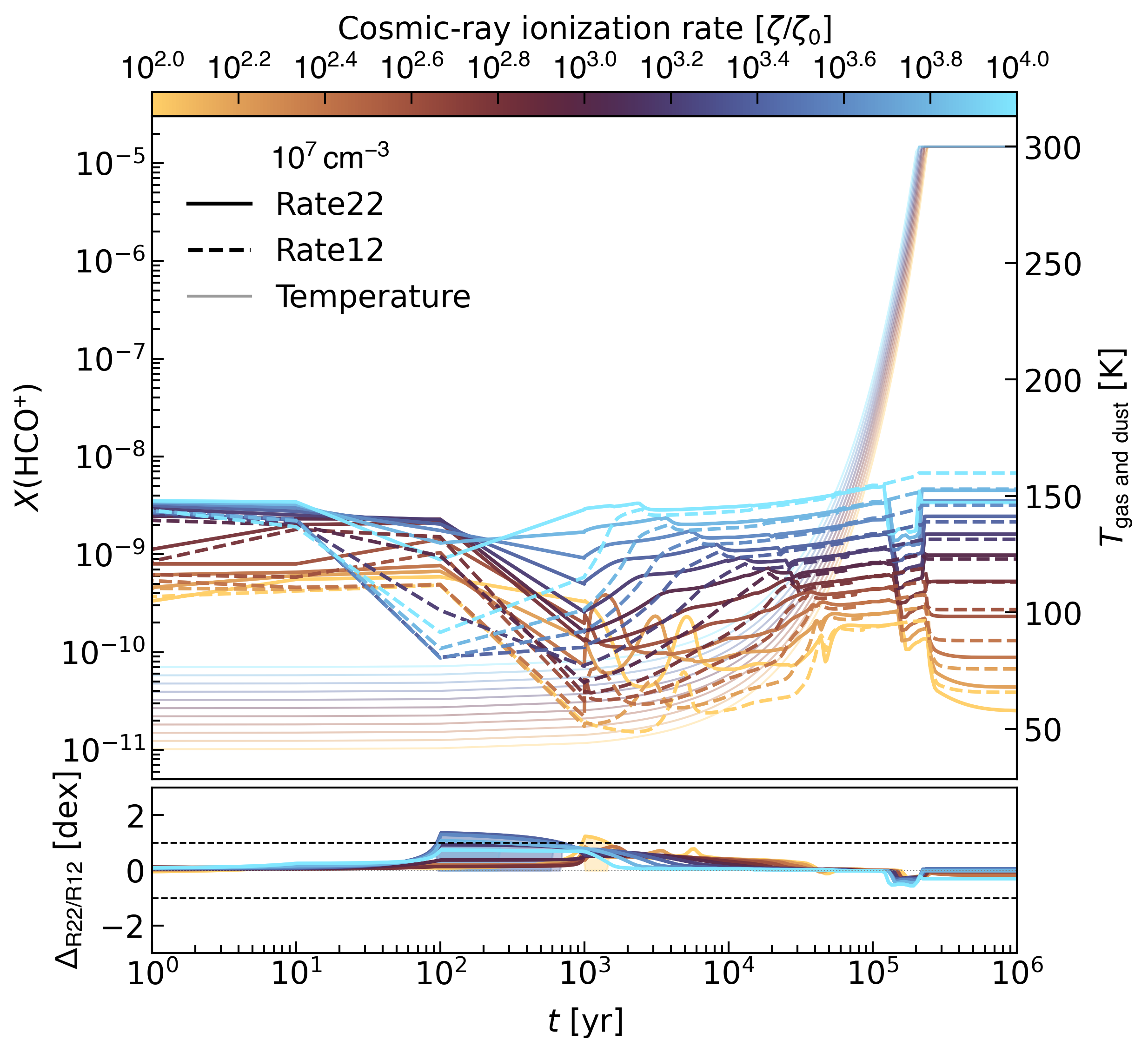}}
    \hfill
    \subfigure{\includegraphics[width=0.33\textwidth]{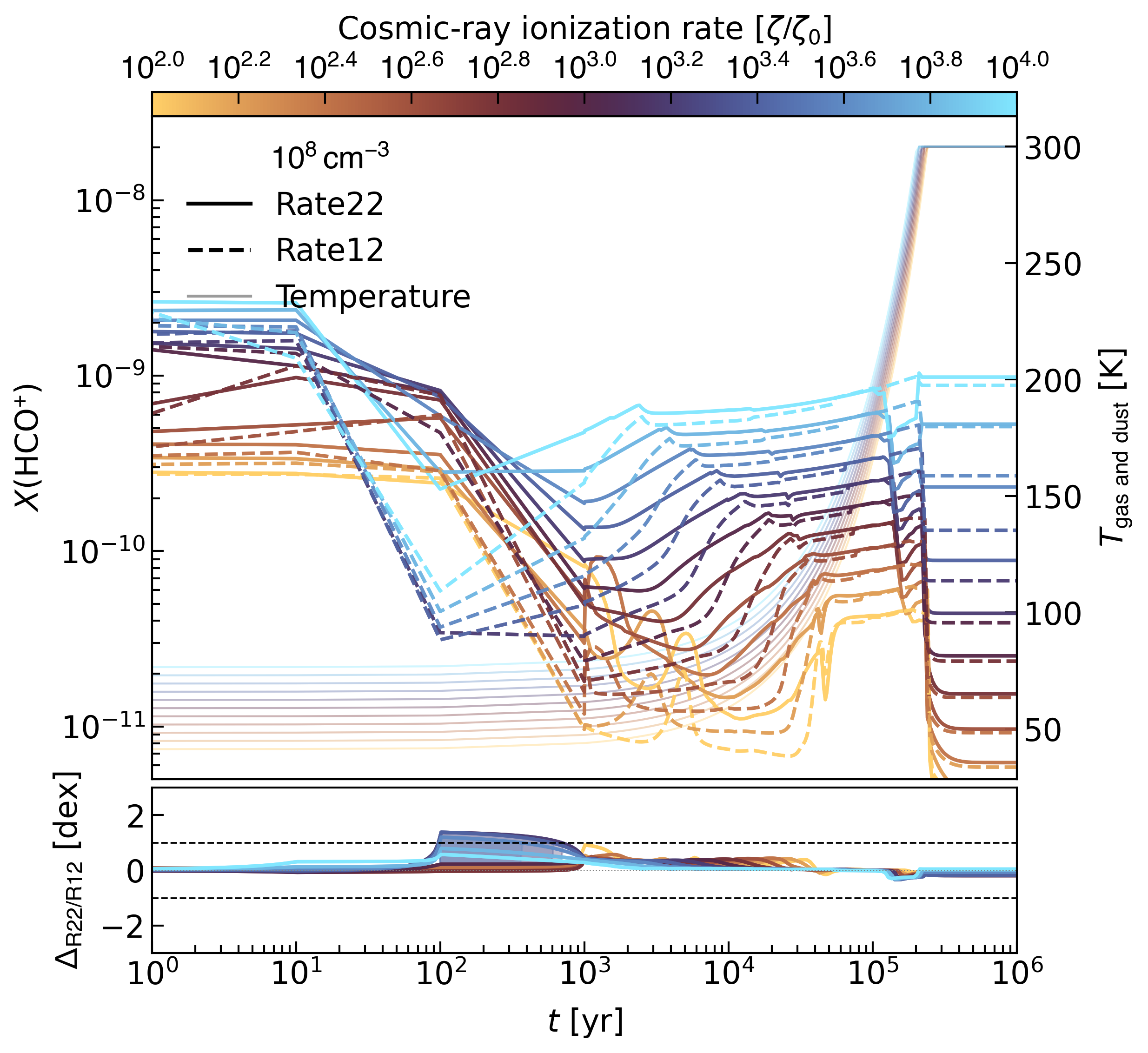}}
    \caption{As in Fig.~\ref{fig:DMG_all}, but for protostellar gas models with $T_{\max}=300$~K and densities $n_{\rm H}=10^6-\dens{8}$. The two networks begin to diverge after $\sim$100~yr, as seen in the bottom panels.}
    \label{fig:PG_300K}
\end{figure*}

\begin{figure*}[ht]
    \centering
    \subfigure{\includegraphics[width=0.33\textwidth]{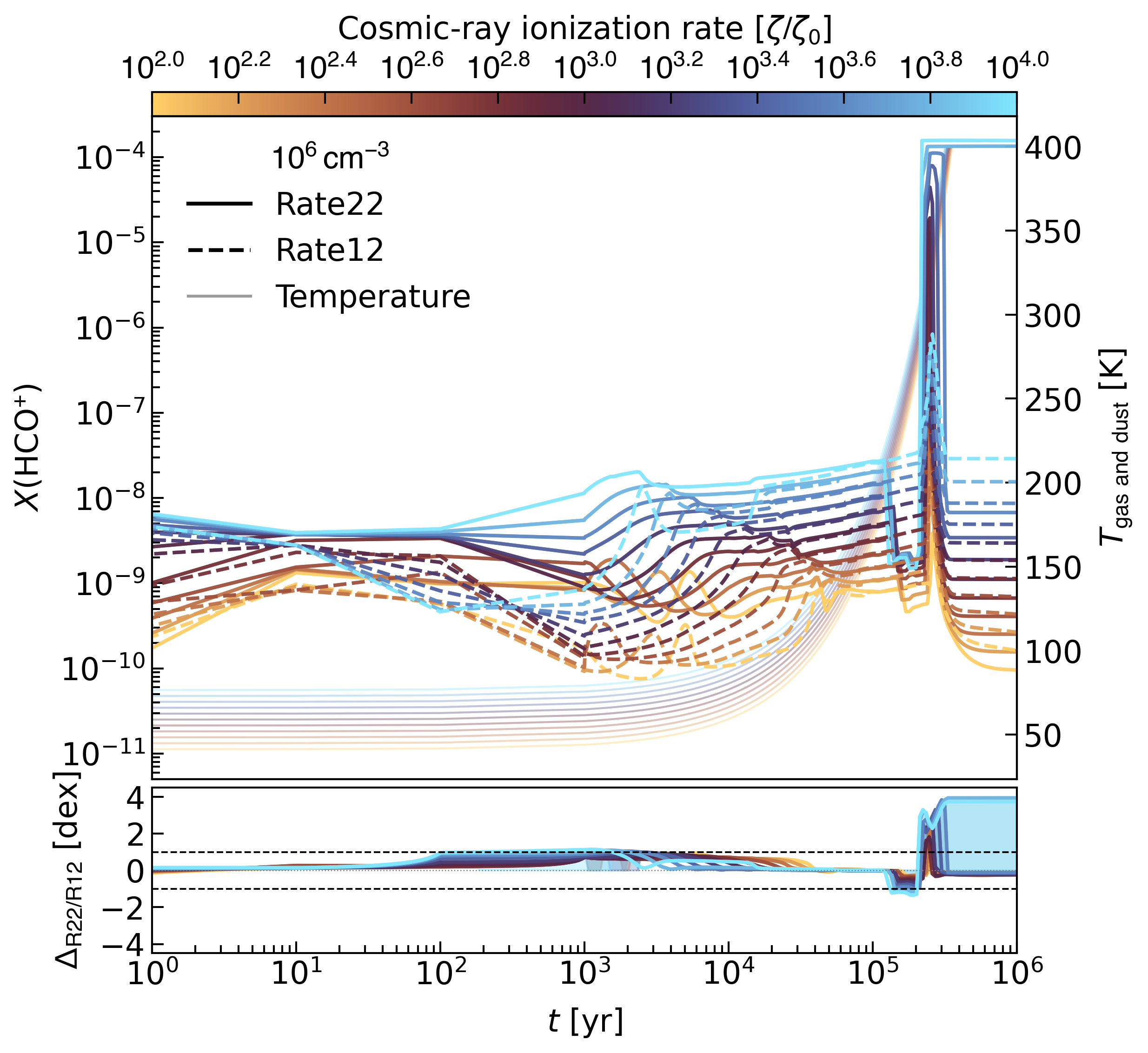}}
    \hfill
    \subfigure{\includegraphics[width=0.33\textwidth]{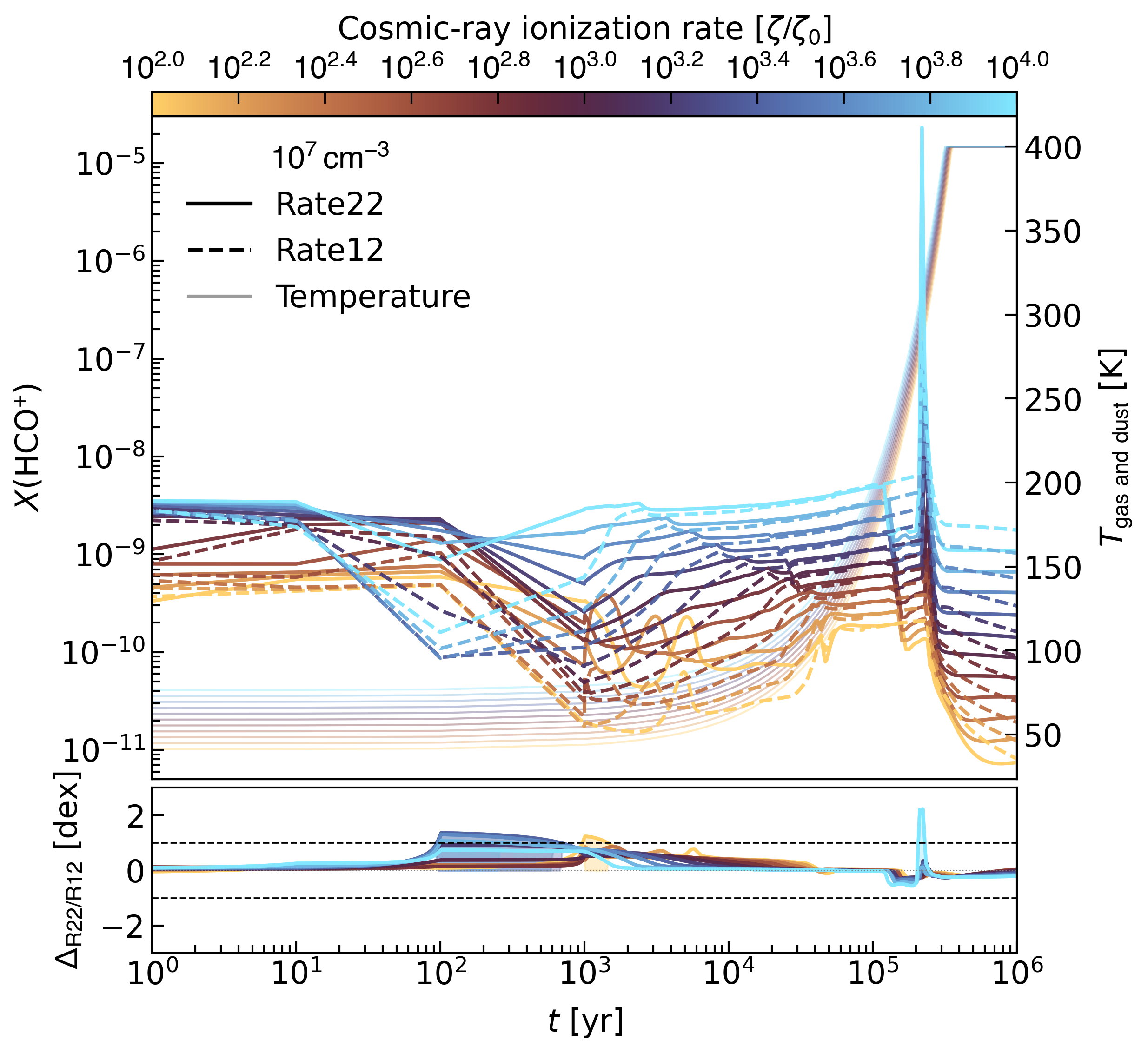}}
    \hfill
    \subfigure{\includegraphics[width=0.33\textwidth]{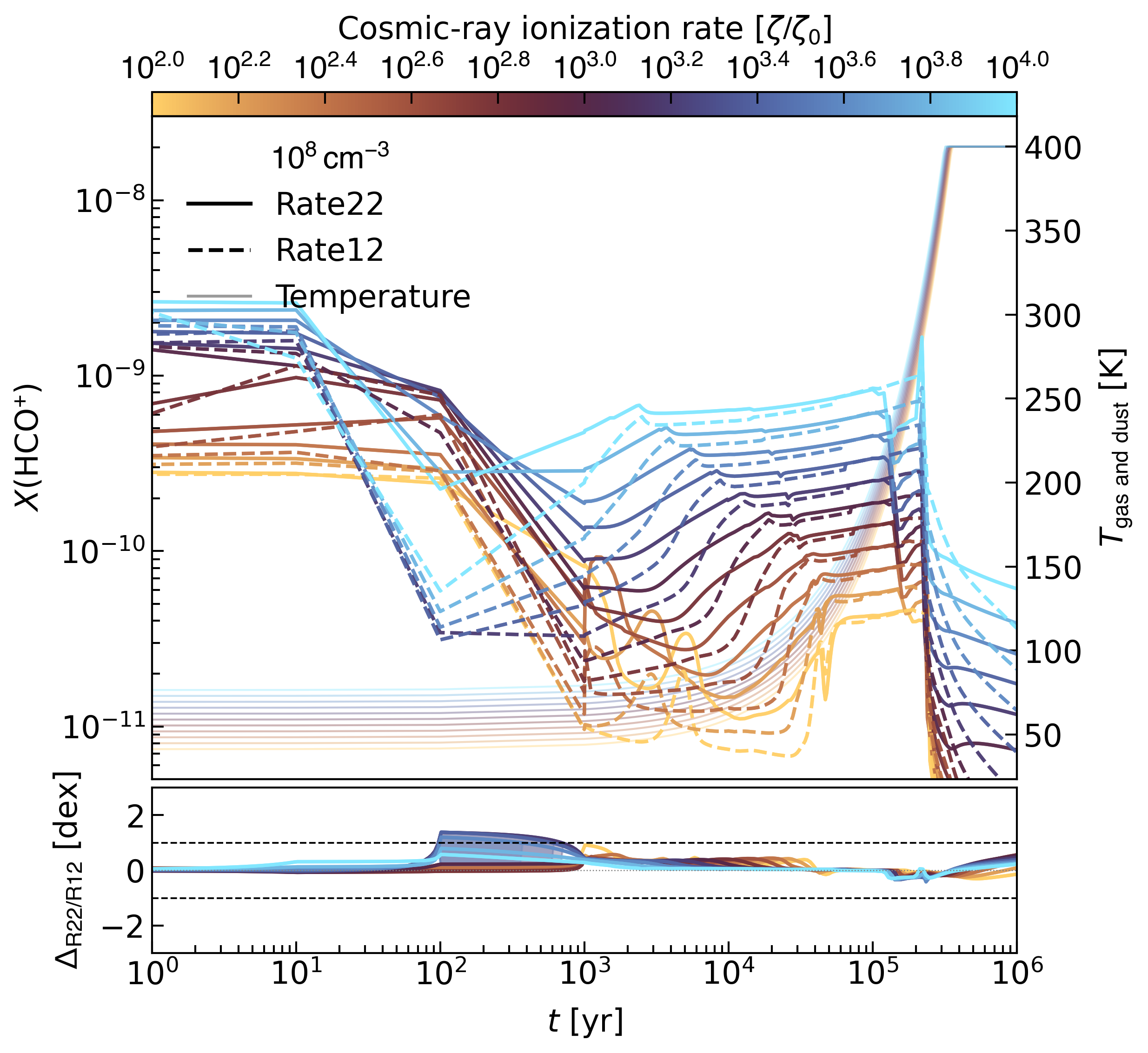}}
    \caption{As in Fig.~\ref{fig:PG_300K}, but for protostellar gas models with $T_{\max}=400$~K.}
    \label{fig:PG_400K}
\end{figure*}

\begin{figure*}[ht]
    \centering
    \subfigure{\includegraphics[width=0.33\textwidth]{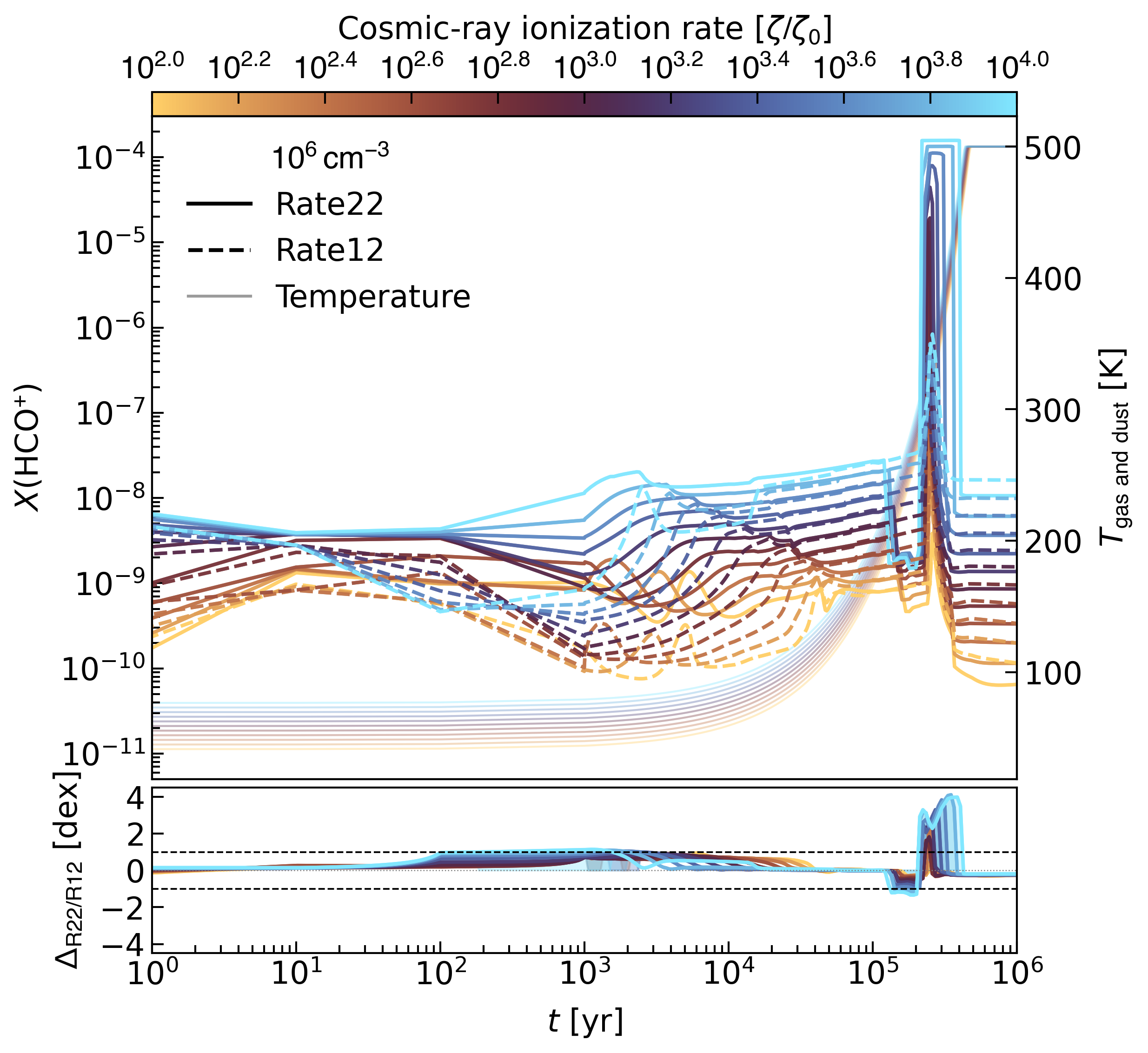}}
    \hfill
    \subfigure{\includegraphics[width=0.33\textwidth]{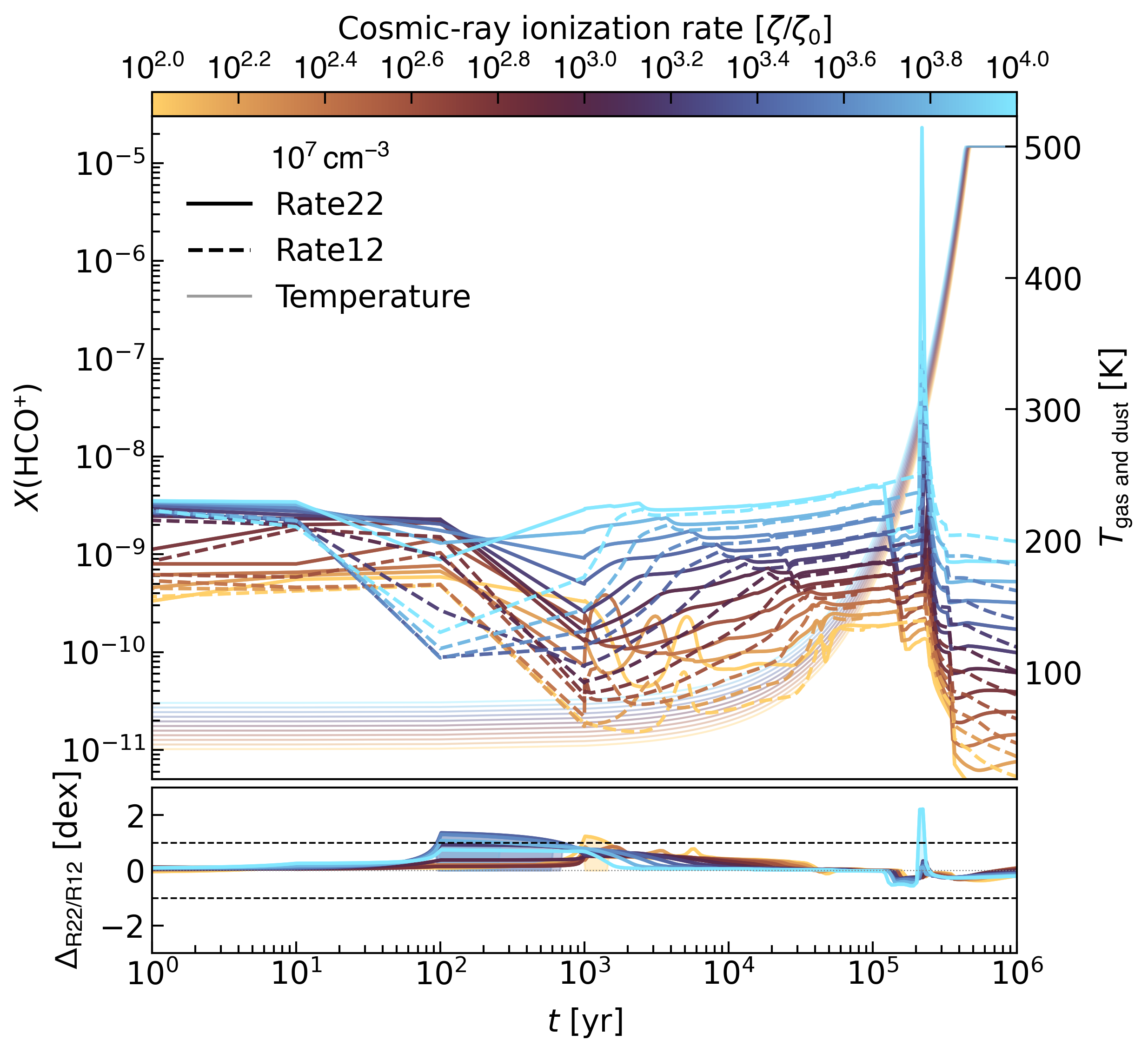}}
    \hfill
    \subfigure{\includegraphics[width=0.33\textwidth]{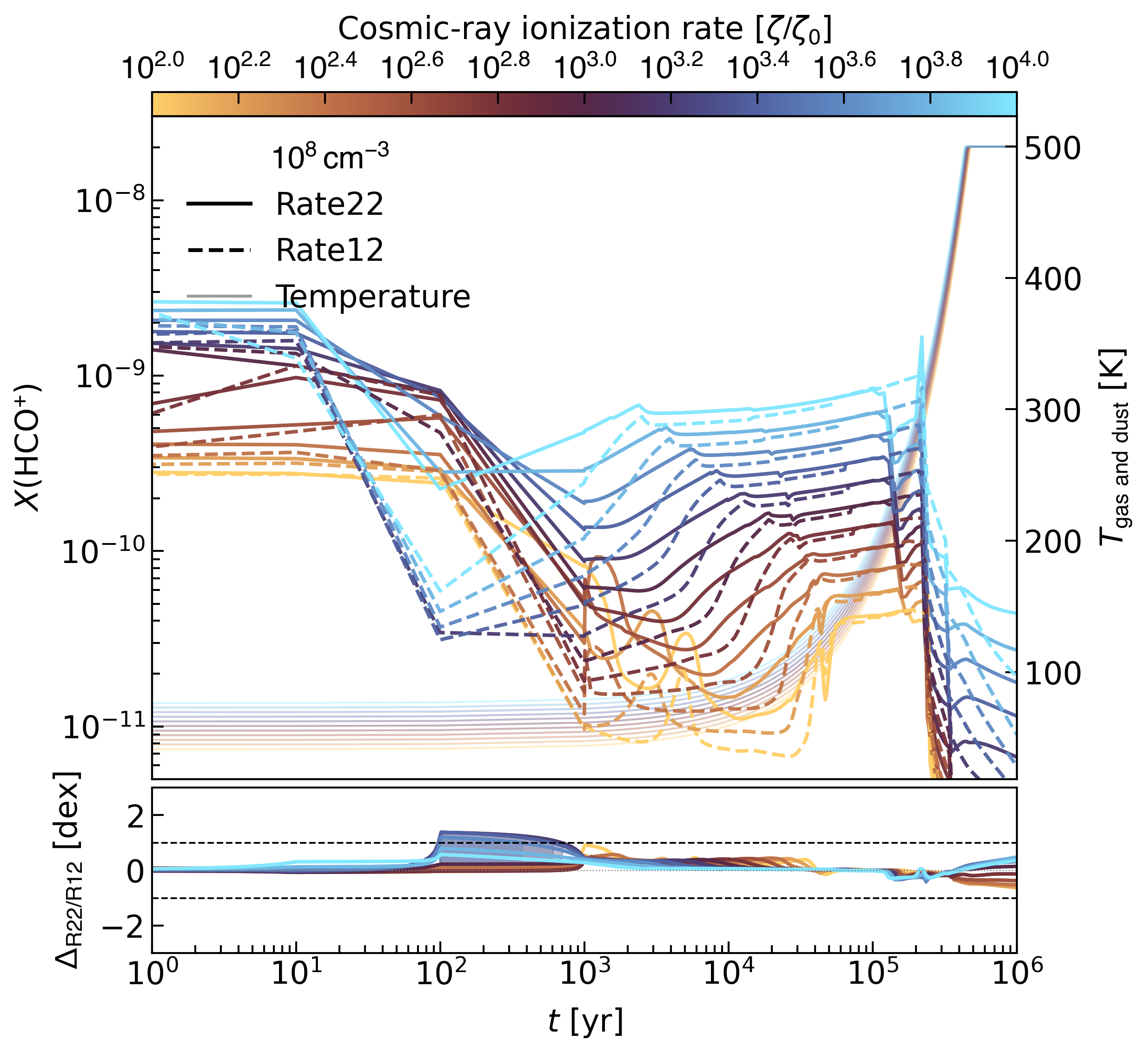}}
    \caption{As in Fig.~\ref{fig:PG_300K}, but for protostellar gas models with $T_{\max}=500$~K.}
    \label{fig:PG_500K}
\end{figure*}

Protostellar gas models cover temperatures of $300-500$~K and densities $n_\mathrm{H} = 10^6-\dens{8}$. The overall chemical evolution and the resulting network differences vary with density, while temperature plays an increasingly important role at late times ($t \gtrsim 10^5$~yr) for $n_\mathrm{H} \leq \dens{7}$ (Fig. \ref{fig:PG_300K}--\ref{fig:PG_500K}). In this regime, the \ch{HCO+} abundance becomes strongly temperature-dependent, with network differences reaching several dex in the most extreme cases.

This is particularly evident at $n_\mathrm{H} = \dens{6}$, where a reversal occurs between 300 and 400~K models for $\zeta > \crir{-14}$. At 300~K, \R{22} predicts substantially lower \ch{HCO+} abundances than \R{12} at $t \gtrsim 10^5$~yr, with differences exceeding 2~dex at the highest ionization rates. At 400~K, the same models show the opposite behavior, with \R{22} producing higher abundances and differences approaching 4~dex.

As in the dense gas models, the networks diverge after $\sim$100~yr. The removal of the \ch{C + HCO+ -> CO + CH+} destruction pathway from \R{22} is again a plausible contributor, particularly in the models producing the highest abundances, although other network differences likely play a role and their relative contributions are difficult to disentangle.

\subsection{Shocked gas}
\label{sec:diff-shocked}

\begin{figure*}[ht]
    \centering
    \subfigure{\includegraphics[width=0.33\textwidth]{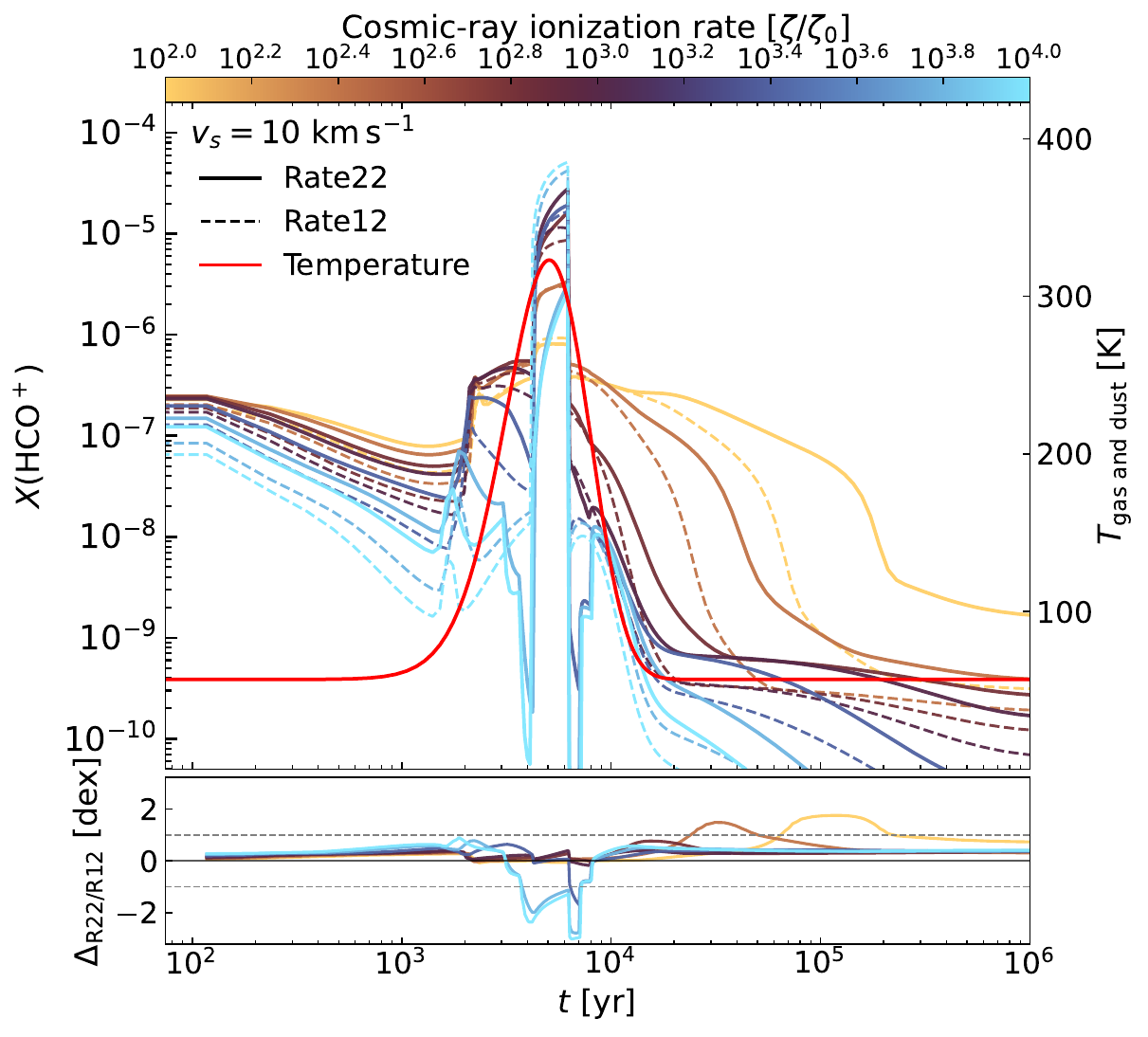}}
    \hfill
    \subfigure{\includegraphics[width=0.33\textwidth]{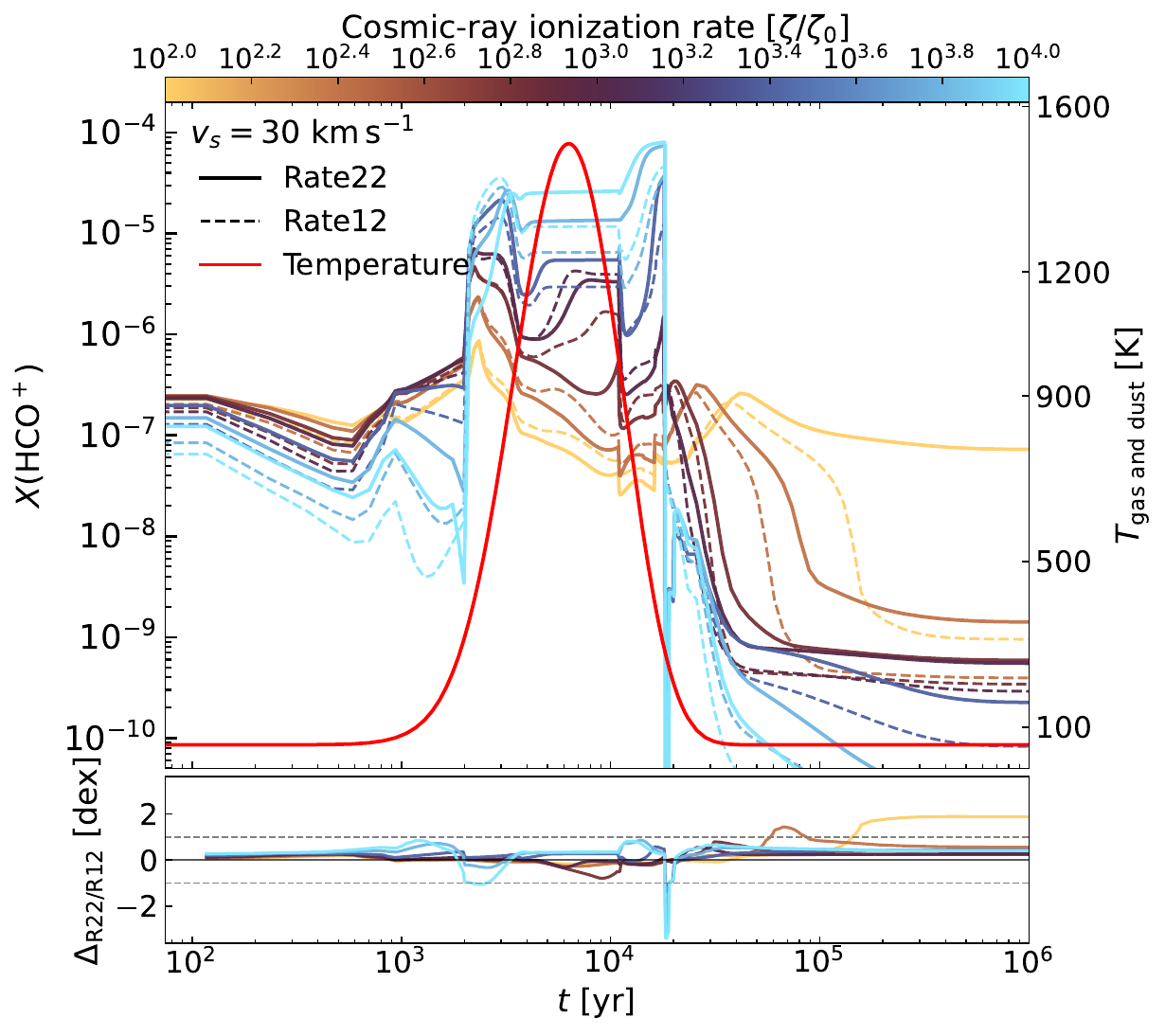}}
    \hfill
    \subfigure{\includegraphics[width=0.33\textwidth]{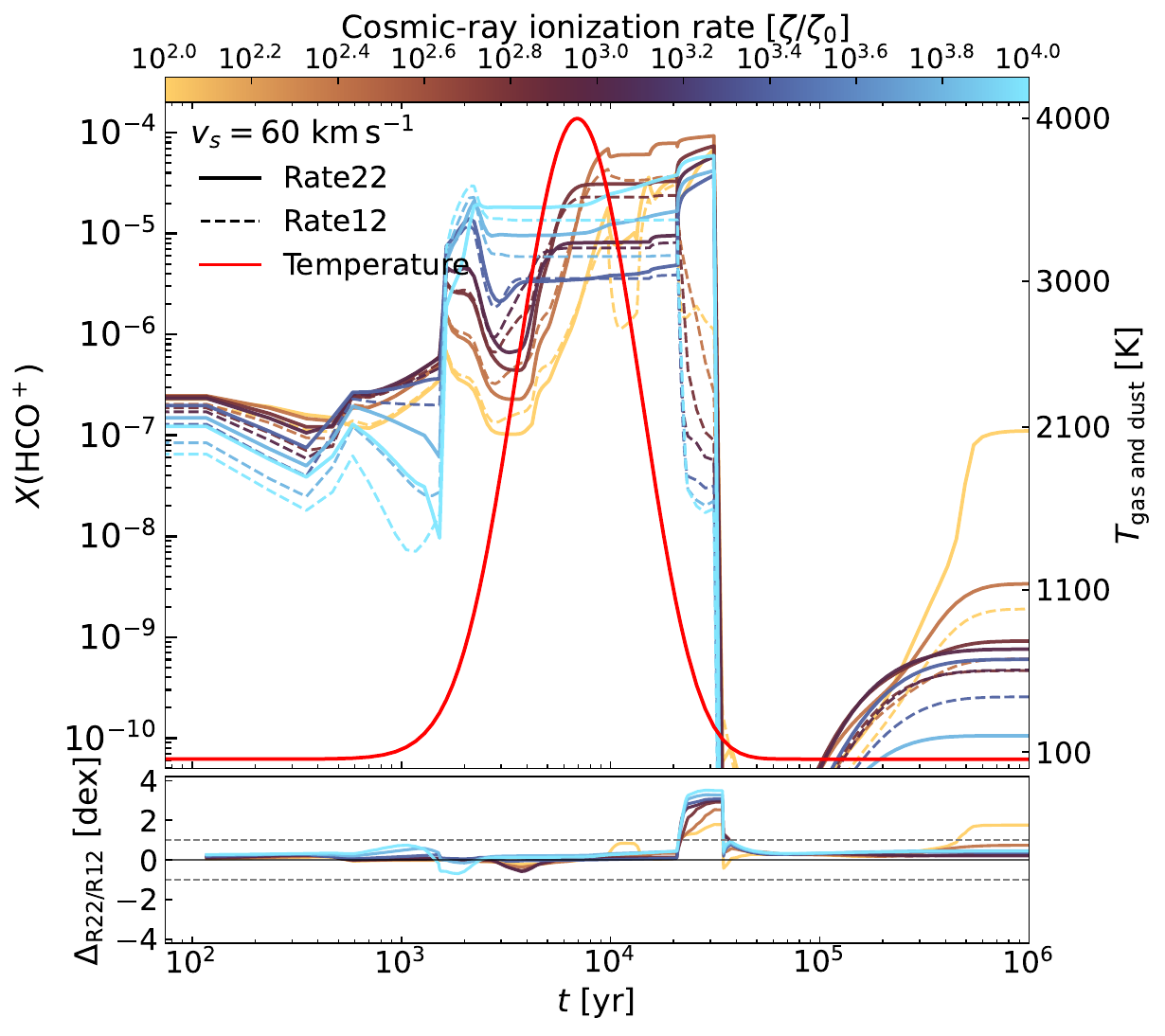}}
    \caption{As in Fig.~\ref{fig:DMG_all}, but for shocked gas models with $n_{\rm pre} = \dens{3}$ for slow, moderate and fast shocks (left to right). The red solid line shows the temperature evolution for a representative model. The largest differences between the networks occur during and shortly after the shock phase, while at late times, as the gas approaches equilibrium, $\Delta_\mathrm{R22/R12}$ tend to stabilize at low values.}
    \label{fig:Shocks_n1e3}
\end{figure*}

\begin{figure*}[ht]
    \centering
    \subfigure{\includegraphics[width=0.33\textwidth]{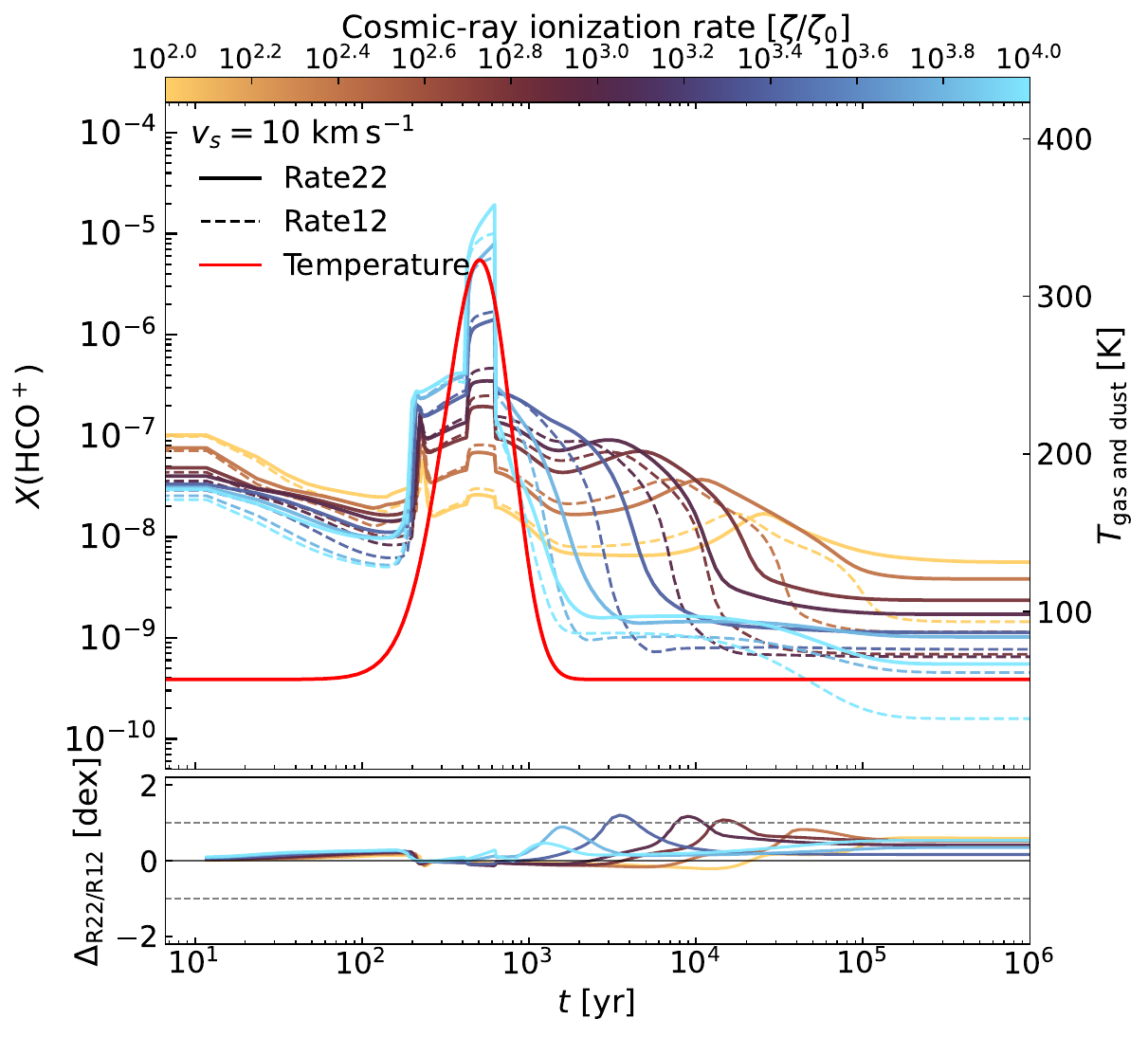}}
    \hfill
    \subfigure{\includegraphics[width=0.33\textwidth]{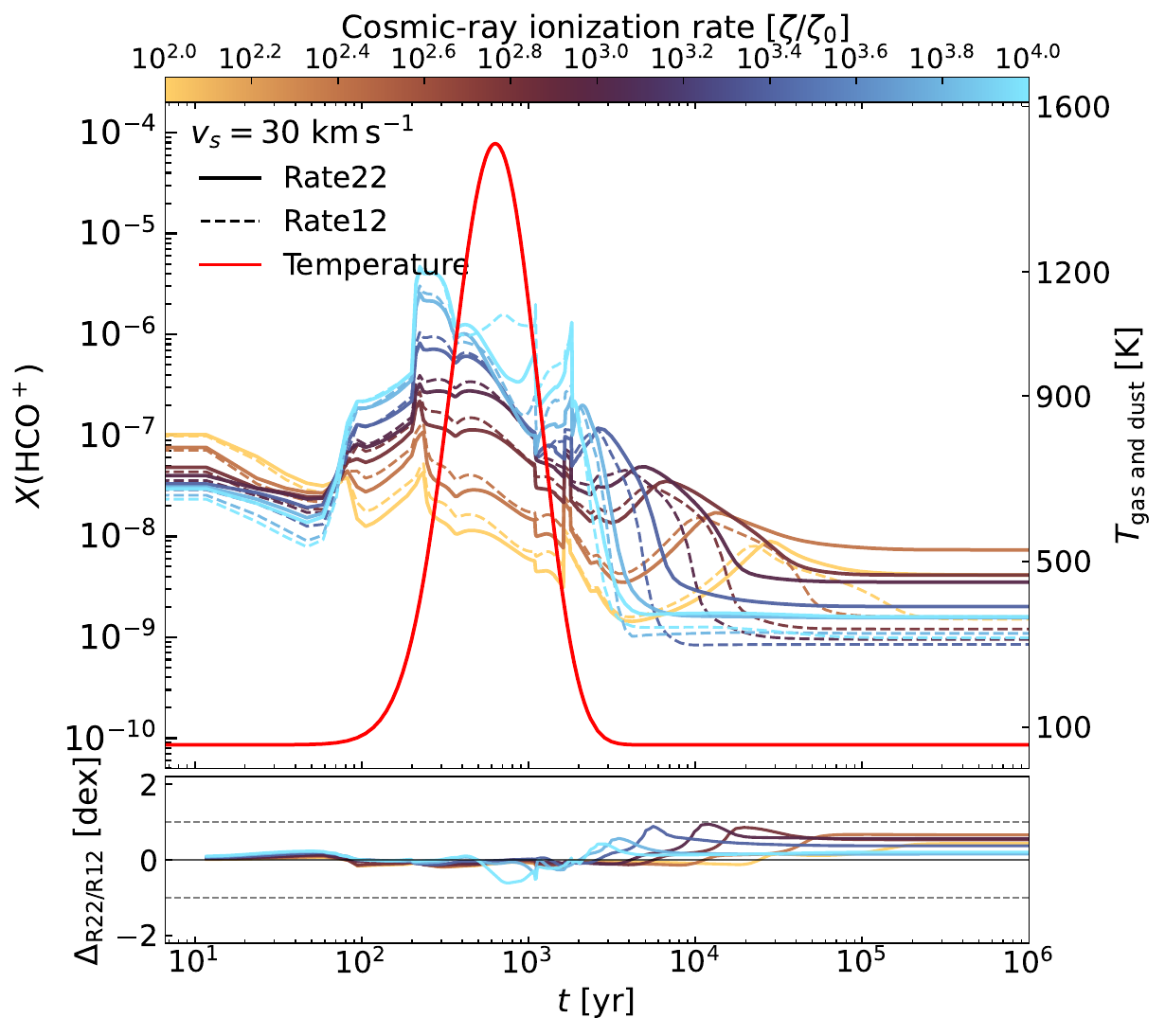}}
    \hfill
    \subfigure{\includegraphics[width=0.33\textwidth]{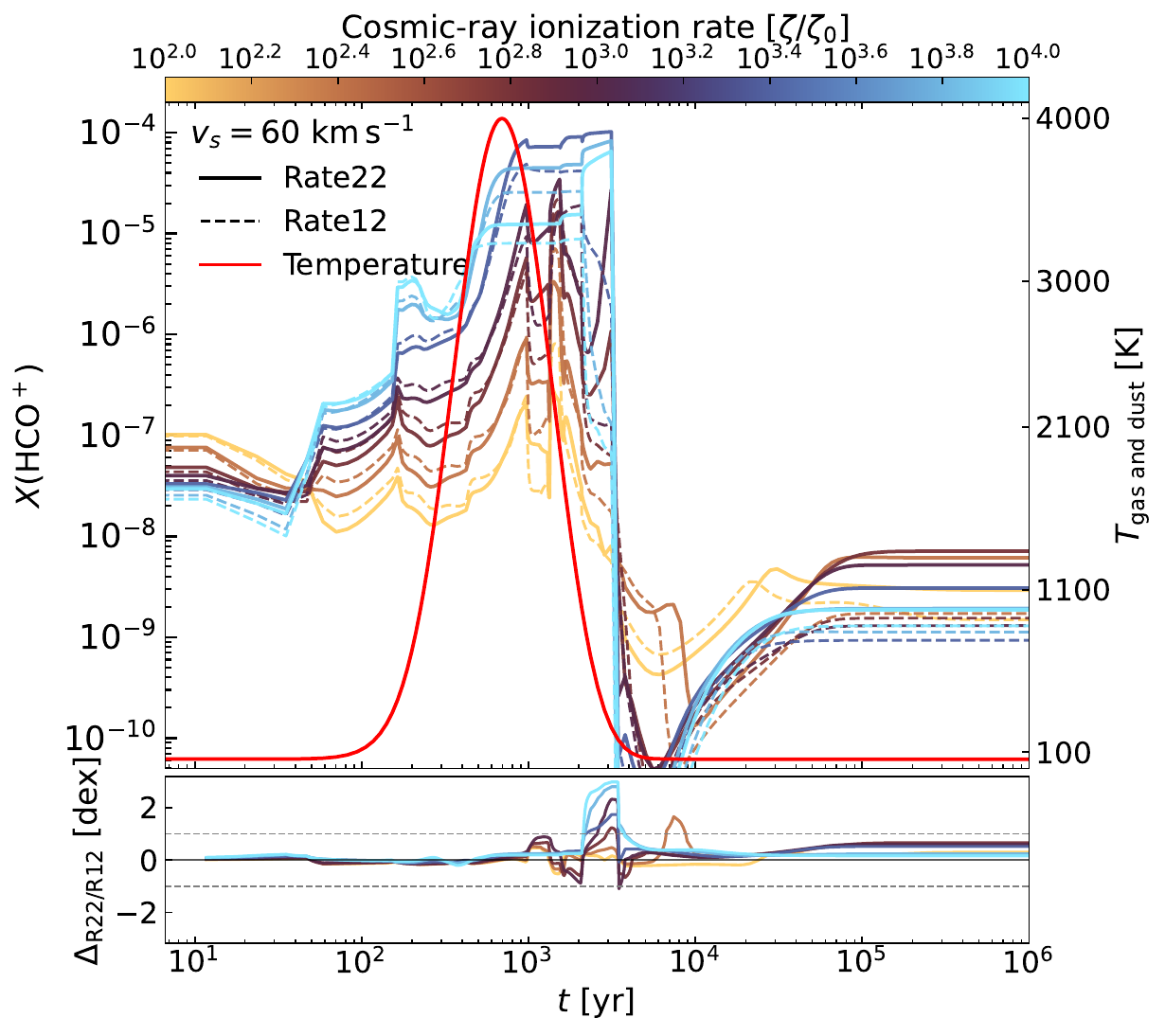}}
    \caption{As in Fig.~\ref{fig:Shocks_n1e3}, but for shocked gas models with $n_{\rm pre} = \dens{4}$.}
    \label{fig:Shocks_n1e4}
\end{figure*}

\begin{figure*}[ht]
    \centering
    \subfigure{\includegraphics[width=0.33\textwidth]{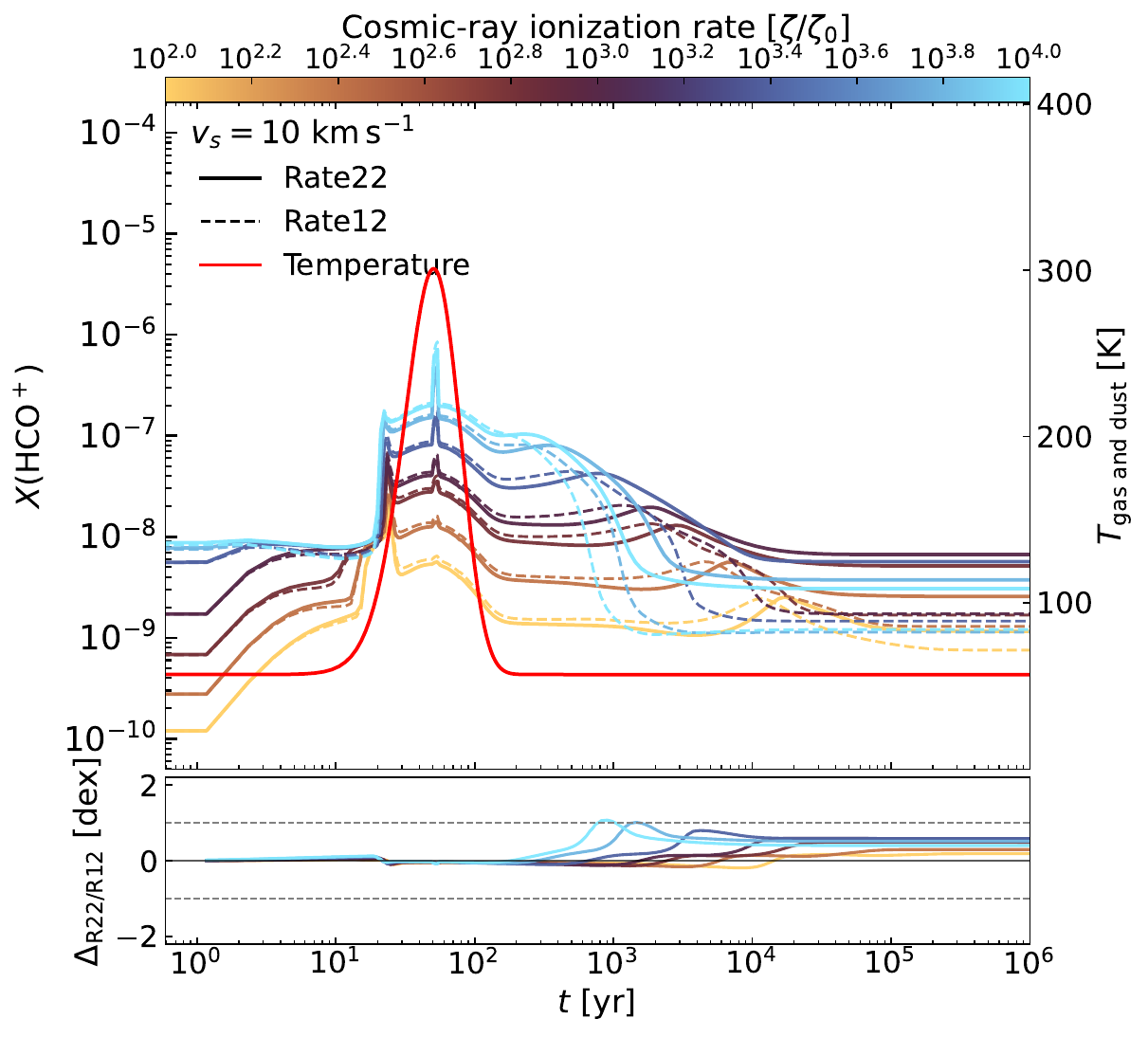}}
    \hfill
    \subfigure{\includegraphics[width=0.33\textwidth]{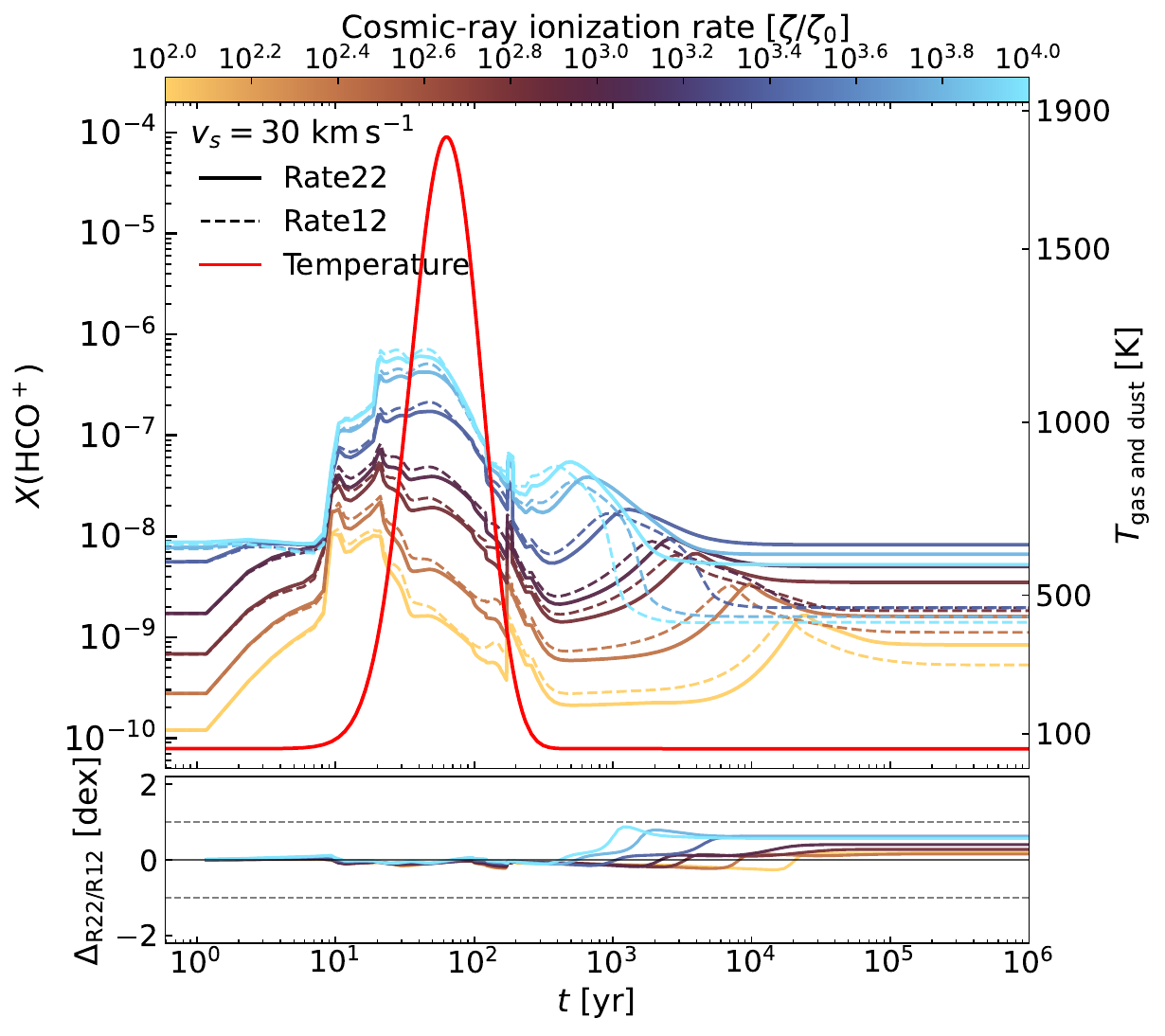}}
    \hfill
    \subfigure{\includegraphics[width=0.33\textwidth]{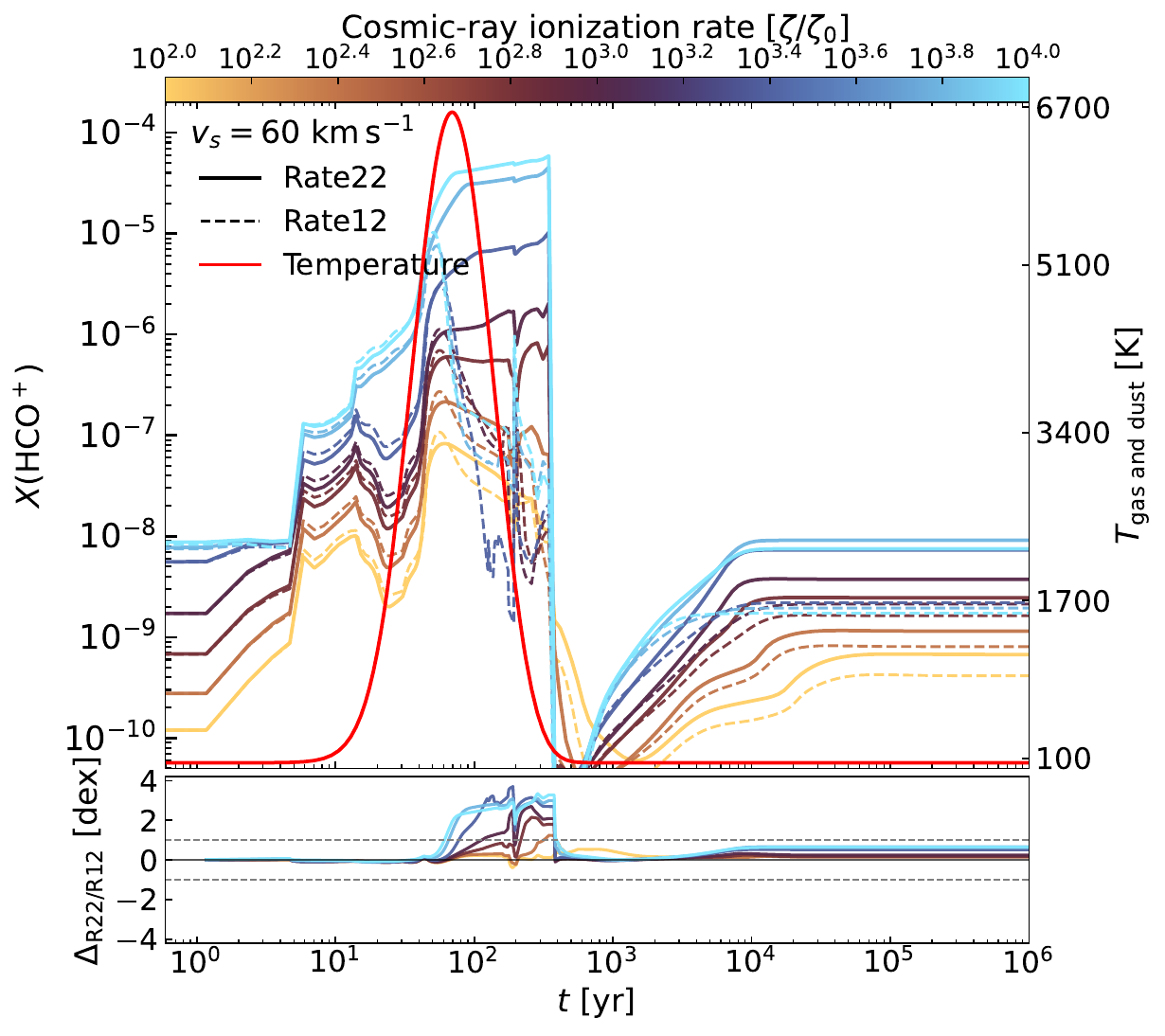}}
    \caption{As in Fig.~\ref{fig:Shocks_n1e3}, but for shocked gas models with $n_{\rm pre} = \dens{5}$.}
    \label{fig:Shocks_n1e5}
\end{figure*}

Shocked gas models cover pre-shock densities of $10^3-\dens{5}$ and a broad range of shock velocities. Here we focus on representative cases of 10, 30, and $\kms{60}$ (Fig.~\ref{fig:Shocks_n1e3}--\ref{fig:Shocks_n1e5}), corresponding to slow, moderate, and fast shocks, respectively. As in the other models, \R{22} tends to produce higher \ch{HCO+} abundances at higher densities. 

At the lowest density, the differences between the two networks depend on shock velocity, ionization rate, and evolutionary stage. In slow and moderate shocks, \R{22} can produce higher \ch{HCO+} abundances during parts of the post-shock phase at low $\zeta$, while at the highest ionization rates \R{12} can exceed \R{22} by more than 2~dex. In contrast, fast shocks produce the strongest \ch{HCO+} enhancements in \R{22} and differences between the networks reaching $\Delta_\mathrm{R22/R12}\geq2$. These large differences are confined to the shock phase and decrease once the system reaches equilibrium, where $\Delta_\mathrm{R22/R12}$ returns to near zero and the \ch{HCO+} abundance becomes primarily governed by the cosmic ray ionization rate.
 
At higher densities, low and moderate shocks produce only minor differences between the networks, while fast shocks again lead to the most pronounced enhancements. The largest difference, 3.73~dex, is found for $n_\mathrm{H}=\dens{5}$ under $\zeta/\zeta_0 = 10^{4}$. This indicates that extremely high \ch{HCO+} abundances arise from the combined effect of high $v_\mathrm{s}$ and high $\zeta$.

\end{appendix}

\end{document}